\documentclass[aps,prd,10pt,notitlepage,superscriptaddress,nofootinbib,numbers]{revtex4-2}

\usepackage[utf8]{inputenc}
\usepackage[T1]{fontenc}
\usepackage{lmodern}
\usepackage{microtype}
\usepackage{amsmath,amssymb,amsfonts,mathrsfs}
\usepackage{bm}
\usepackage{graphicx}
\usepackage{float}
\usepackage[dvipsnames]{xcolor}
\usepackage{hyperref}
\hypersetup{colorlinks=true,citecolor=Purple,linkcolor=Purple,urlcolor=Purple,breaklinks=true}
\usepackage[capitalize]{cleveref}
\usepackage{tikz}

\newcommand{\Del}{\nabla}
\newcommand{\dd}{\mathrm{d}}
\newcommand{\Fh}{\,{}_2F_1}

\definecolor{lime}{HTML}{A6CE39}
\DeclareRobustCommand{\orcidicon}{%
	\begin{tikzpicture}
		\draw[lime, fill=lime] (0,0)
		circle [radius=0.16]
		node[white] {{\fontfamily{qag}\selectfont \tiny ID}};
		\draw[white, fill=white] (-0.0625,0.095)
		circle [radius=0.007];
	\end{tikzpicture}
	\hspace{-2mm}
}

\foreach \x in {A, ..., Z}{%
	\expandafter\xdef\csname orcid\x\endcsname{\noexpand\href{https://orcid.org/\csname orcidauthor\x\endcsname}{\noexpand\orcidicon}}
}

\newcommand\orcidZijun{{\href{https://orcid.org/0009-0001-1322-6303}{\orcidicon}}}
\newcommand\orcidJonathan{{\href{https://orcid.org/0000-0001-9291-0893}{\orcidicon}}}
\newcommand\orcidEdson{{\href{https://orcid.org/0000-0001-9929-5977}{\orcidicon}}}
\newcommand\orcidMinyong{{\href{https://orcid.org/0000-0001-5577-575X}{\orcidicon}}}

\begin{document}

\title{Symmetric Thin-Shell Wormholes from a Regularized String-Cloud Black Hole: Energy Conditions, Multi-Fluid Stability, and Optical Signatures}

\author{Zijun Zhong\orcidZijun\!\!}
\email{zjzhong@mail.bnu.edu.cn}
\affiliation{School of Physics and Astronomy, Beijing Normal University, Beijing 100875, China}

\author{Jonathan A. Rebou\c{c}as\orcidJonathan\!\!}
\email{jalvesreboucas@ifce.edu.br}
\affiliation{Instituto Federal de Educa\c{c}\~ao, Ci\^encia e Tecnologia do Cear\'a (IFCE), Iguatu, Brazil}

\author{Edson Otoniel\orcidEdson\!\!}
\email{cedson.otoniel@ufca.edu.br}
\affiliation{Universidade Federal do Cariri (UFCA), Instituto de Forma\c{c}\~ao de Educadores--IFE, R. Oleg\'ario Emidio de Araujo S/N, Brejo Santo--CE, 63.260-000, Brazil}

\author{Minyong Guo\orcidMinyong\!\!}
\email{minyongguo@bnu.edu.cn}
\affiliation{School of Physics and Astronomy, Beijing Normal University, Beijing 100875, China}
\affiliation{Key Laboratory of Multiscale Spin Physics (Beijing Normal University), Ministry of Education, Beijing 100875, China}

\date{\today}

\begin{abstract}
We construct a symmetric thin-shell wormhole from two copies of the positive-lapse exterior of a rationally regularized Letelier--Alencar string-cloud black hole with a finite-curvature anti--de Sitter core and string-cloud asymptotics. Using the Darmois--Israel formalism, we derive the surface stresses, energy-condition combinations, conservation law, and effective potential. Every admissible static throat has negative surface energy density, so the weak and dominant energy conditions are violated, whereas the null condition restricted to directions tangent to the shell, and consequently the intrinsic strong condition, holds only in the inner sampled exterior. The radial-vacuum relation of the bulk source removes the external momentum flux. We derive a general local-stability criterion for density- and radius-dependent equations of state and apply it to five surface-fluid models, calibrating one parameter at each static throat. Across the sampled families, barotropic shells remain unstable, whereas the other closures can admit parameter-dependent stable branches. An exact unregularized control shows that the energy-condition structure and the radial-stability signs of the first three closures persist without the rational regulator, whereas the cosmic-Chaplygin stability sign can change while the selected calibration branch remains real and simple. Backward ray tracing shows that throat location primarily reorganizes higher-order and through-throat images, while string-cloud strength and inclination control annular structure and Doppler asymmetry. Comparisons of one- and two-sided source configurations isolate genuine far-side images and show that the central dark region depends on the source distribution.
\end{abstract}

\maketitle

\tableofcontents

\section{Introduction}\label{sec:introduction}

Traversable wormholes provide a particularly sharp arena for testing how spacetime topology, causal structure, and the material content of gravitation constrain one another. In the Morris--Thorne framework, a throat connecting two regions can remain horizon free and traversable only if its geometry satisfies a flare-out condition whose realization in general relativity is tied to violations of the classical energy conditions \cite{MorrisThorne1988,MorrisThorneYurtsever1988,HochbergVisser1998}. This local requirement is reinforced by global results such as topological censorship, which excludes observable nontrivial topology under appropriate causal, asymptotic, and averaged-null-energy assumptions \cite{FriedmanSchleichWitt1993}. Quantum field theory does permit negative-energy densities in restricted circumstances, but quantum inequalities severely constrain their magnitude, duration, and spatial distribution \cite{FordRoman1996}. Consequently, proposals that reduce the integrated amount of exotic matter must still be distinguished from constructions that genuinely evade local or quantum restrictions \cite{VisserKarDadhich2003,FewsterRoman2005}. These considerations do not remove wormholes from theoretical study; rather, they make the localization, interpretation, and dynamical response of the supporting source central questions.

Thin-shell wormholes address the localization problem by replacing a smooth exotic-matter distribution with a distributional source confined to a timelike junction. The mathematical basis is the junction formalism for non-null hypersurfaces, in which the jump of the extrinsic curvature determines the surface stress tensor \cite{Israel1966}. Visser's surgical construction showed that two exterior regions can be glued at a spherical boundary to produce a traversable geometry whose nontrivial matter content is concentrated on the throat \cite{Visser1989}. Poisson and Visser subsequently cast radial perturbations of a static shell as motion in an effective potential and obtained the standard linearized stability criterion \cite{PoissonVisser1995}. This framework has since been generalized to arbitrary spherically symmetric seed metrics, asymmetric junctions, momentum fluxes, charged backgrounds, cosmological terms, higher dimensions, and different throat topologies \cite{Eiroa2008SphericalStability,GarciaLoboVisser2012,EiroaRomero2004Charged,LoboCrawford2004,DiasLemos2010}. The geometric interpretation of flare-out also requires care for a distributional throat, because conditions derived for smooth wormhole geometries cannot always be transferred unchanged to a thin shell \cite{MazharimousaviHalilsoy2014FlareOut}. Reviews of the subject emphasize that the seed geometry and the shell equation of state are equally important: changing either one can alter the admissible static domain and the sign of the second derivative of the effective potential \cite{KokubuHarada2020,RadhakrishnanEtAl2024Review}.

Regular black holes supply an especially useful class of seed spacetimes because they replace the central curvature singularity by a finite-curvature core while retaining horizon structures that can be compared with their singular counterparts. Modern exact examples arise from nonlinear electrodynamics, effective matter distributions, limiting-curvature constructions, and running-coupling improvements \cite{AyonBeatoGarcia1998,AyonBeatoGarcia2005,Bronnikov2001RegularMagnetic,Dymnikova2004RegularCharged,Hayward2006,Frolov2016,IshibashiOhtaYamaguchi2021}. Their sources commonly satisfy the radial-vacuum relation $T^t{}_t=T^r{}_r$, although the remaining energy conditions depend on the detailed anisotropy and regularization mechanism \cite{Zaslavskii2010,BalartVagenas2014}. Regularity by itself therefore does not determine the properties of a thin-shell wormhole constructed from the exterior: the junction excises the core, while the lapse function inherited from the regular geometry controls the surface stresses and radial dynamics. This separation has motivated thin-shell studies based on Hayward and other regular geometries, black-bounce spacetimes, and quantum-corrected backgrounds \cite{HalilsoyOvgunMazharimousavi2014,SharifMumtaz2016Hayward,LoboSimpsonVisser2020,JavedEtAl2023QuantumCorrected,Javed2024}. Related work in massive, scalar--tensor, and nonminimally coupled theories further shows that conclusions about exoticity are theory dependent; surface matter satisfying standard energy conditions can occur when the junction equations differ from those of ordinary Einstein gravity \cite{TangphatiEtAl2020dRGT,ForghaniEtAl2020EnergyConditions,EiroaEtAl2026ScalarTensor}.

A distinct line of development concerns black holes sourced by clouds or fluids of strings. Letelier formulated the covariant stress tensor of a continuous distribution of one-dimensional strings and obtained a Schwarzschild-like exterior modified by a constant string contribution \cite{Letelier1979}. The model has since been extended to more general string fluids with nonconstant constitutive behavior and has been used to study horizons, thermodynamics, geodesics, and shadows \cite{Santos2025StringFluids}. Alencar, Landim, and Costa Filho introduced an additional magnetic-like component in the string bivector, producing a hypergeometric correction to the lapse while recovering the Letelier regime in the appropriate limit \cite{Alencar2025}. The resulting geometry remains singular at the center. Muniz \textit{et al.} then applied a rational Dagum-type regulator to this Letelier--Alencar source and obtained an exact black-hole solution with finite curvature invariants, a locally anti--de Sitter core, and a string-cloud exterior \cite{Muniz2026}. That work mapped the horizon sector and examined energy conditions, thermodynamics, and shadow observables, thereby providing both the direct seed metric and the physical parameter domain used in the present construction. The appearance of an anti--de Sitter core also places this solution within the wider family of nonsingular geometries whose central vacuum structure differs from the more common de Sitter core.

String matter has already appeared in several thin-shell settings, but the existing constructions address different seed geometries. Richarte and Simeone built thin-shell wormholes directly from the singular Letelier string-cloud spacetime and analyzed the amount of exotic matter, traversability, and radial stability \cite{RicharteSimeone2008}. More recently, Bardeen--AdS thin-shell wormholes surrounded by a fluid of strings were studied with barotropic, variable phantom-like, and variable Chaplygin closures \cite{WaseemEtAl2025StringFluid}, while Hayward--Letelier--AdS geometries were used to examine the influence of the regularization, string-cloud, and anti--de Sitter parameters on similar variable equations of state \cite{HouedokouEtAl2026HaywardLetelier}. These studies demonstrate that string contributions and regular black-hole parameters can materially change the stability landscape. They do not, however, treat the rationally regularized Letelier--Alencar geometry, whose hypergeometric string correction, anti--de Sitter core, and non-Minkowskian string-cloud exterior constitute a different background. Wormholes supported by nonstandard matter profiles and regularizing corrections in other contexts provide additional motivation for isolating how such a seed geometry affects energy conditions and equilibrium \cite{ReboucasEtAl2026QuantumImproved}.

The shell equation of state supplies the second ingredient of the stability problem. A linear barotropic law is the simplest local closure and remains useful for exposing the purely geometrical part of the perturbation criterion. Phantom-like fluids, originally introduced in cosmology through equations of state with parameter below $-1$, have also been adopted as phenomenological wormhole sources \cite{GonzalezDiaz2003,Sushkov2005Phantom}. For thin shells, phantom-like relations were studied in constant and radius-dependent forms \cite{Kuhfittig2010}, and variable equations of state were shown to add an explicit radial derivative to the effective-potential criterion \cite{Varela2015}. Chaplygin-type laws provide a broader nonlinear family. The ordinary and generalized Chaplygin gases have been applied to spherical thin-shell wormholes in several black-hole backgrounds \cite{EiroaSimeone2007,Eiroa2009GCG,EiroaFigueroaAguirre2012BornInfeld}. Generalized cosmic and modified cosmic Chaplygin models introduce further nonlinear parameters and have been investigated in charged and cosmological thin-shell geometries \cite{SharifAzam2013GCCG,SharifMumtaz2014}. The modified cosmic form used below is taken as a surface closure only; its functional ancestry in interacting cosmological fluids does not imply that viscosity, scale-factor dependence, or a volumetric cosmological interpretation is inherited by the shell \cite{SadeghiFarahani2013}. The polymer-black-hole analysis of Javed \textit{et al.} is a particularly close methodological comparison because it treats barotropic, variable phantom, and variable Chaplygin laws within the same symmetric thin-shell problem \cite{Javed2024}.

Beyond the standard requirement that the effective potential have positive curvature at equilibrium, recent studies have clarified complementary aspects of thin-shell stability. Thermodynamic stability is logically distinct from radial mechanical stability, although both can be imposed on the same equilibrium family \cite{ForghaniEtAl2019Thermodynamic,EiroaEtAl2024ChargedThermo}. For barotropic shells with vanishing momentum flux, the radial-stability criterion can be related to photon and antiphoton spheres of the seed spacetime \cite{TsukamotoKokubu2024PhotonOrbits}. Conversely, under the assumptions of an asymptotically flat seed and a barotropic surface fluid, an equilibrium throat placed at the innermost photon sphere is generically unstable \cite{Mazharimousavi2024InnermostPhotonSphere}. These results underscore the need to state the asymptotic structure, flux term, and equation-of-state assumptions explicitly before transferring a stability conclusion from one background to another.

Against this dynamical background, the optical appearance provides a complementary probe of how the same junction reorganizes null geodesics. Although wormholes remain hypothetical, horizon-scale very-long-baseline interferometry at $1.3\,\mathrm{mm}$ has reconstructed the compact, gravitationally lensed emission surrounding M87* and Sgr~A*, making the near-critical image morphology of supermassive compact objects an important observational probe of strong gravitational fields \cite{theeventhorizontelescopecollaborationFirstM87Event2019,collaborationFirstSagittariusEvent2022}. These measurements are not direct optical images of event horizons; rather, they resolve synchrotron emission on horizon scales. Theoretically, such images have been studied mainly through two complementary approaches: general-relativistic ray tracing of radiation propagation and analyses of features such as the shadow, lensing ring, and photon ring as probes of strong-field spacetime geometry \cite{luminetImageSphericalBlack1979a,vincentGYOTONewGeneral2011,grallaBlackHoleShadows2019,grallaShapeBlackHole2020}. Previous studies have shown that image morphology depends not only on the propagation geometry but also on the emission model, electrodynamic corrections, magnetic fields, accretion-flow structure, and boundary or junction conditions \cite{huQEDEffectBlack2021a,zengQEDAccretionFlow2022,houImageKerrMelvinBlack2022,zhangImagingThickAccretion2024a,caoImagingSignaturesIsrael2026,wangHowDifferentAre2023}. Consequently, the observed shadow and ring structures generally arise from the combined effects of spacetime geometry and the emitting environment. Although a wormhole can locally possess an exterior geometry closely resembling that of a black hole, it changes the global continuation of null geodesics and may therefore produce distinct imaging signatures. Previous studies have found that the wormhole topology and the symmetry between its two exterior regions can affect the shadow boundary and ring structure and may introduce additional image components. For reflection-symmetric or two-ended wormholes, rays traversing the throat can carry radiation from emitters in the other exterior to the observer's side, producing additional images or light rings while preserving an overall black-hole-like appearance \cite{ohgamiWormholeShadows2015,paulObservationalSignaturesWormholes2020,olmoNewLightRings2022,ruedaTraversableWormholesMultiple2025}. For reflection-asymmetric thin-shell wormholes, the inequivalent photon effective potentials on the two sides can allow the far-side photon sphere to contribute to the shadow edge and can generate double shadows, additional photon rings, or lensing bands \cite{wielgusReflectionasymmetricWormholesTheir2020,wangNovelShadowsAsymmetric2020,pengObservationalSignatureAdditional2021a,guerreroDoubleShadowsReflectionasymmetric2021,Tsukamoto2021DoubleShadows,heObservationalAppearanceExtra2024}. This optical context complements the energy-condition and stability questions above and leads to the combined problem addressed here.

The literature therefore leaves a specific problem open: how a symmetric thin-shell wormhole behaves when its seed is the rationally regularized Letelier--Alencar black hole, and how its surface energy conditions and radial stability respond to several inequivalent shell fluids within one consistent normalization, together with how the string-cloud strength and throat location affect its optical appearance. The present work addresses this problem in four-dimensional general relativity. Two identical copies of the connected positive-lapse exterior are joined at a timelike spherical hypersurface outside the outer event horizon. We derive the dynamical surface density and tangential pressure from the Darmois--Israel conditions, distinguish energy conditions intrinsic to the $2+1$-dimensional shell from their four-dimensional distributional counterparts, and exploit the bulk relation $T^t{}_t=T^r{}_r$ to show that the symmetric shell has no external momentum-flux term. The throat dynamics is written in terms of an effective potential, and we derive a general local-stability criterion for surface equations of state with both density dependence and explicit radial dependence.

We then specialize the general criterion to five surface-fluid models: a linear barotropic law, variable phantom and variable Chaplygin equations of state, a generalized cosmic Chaplygin gas, and a modified cosmic Chaplygin gas. In each case, one natural equation-of-state parameter is fixed by the static junction condition at the selected radius and is held fixed during the perturbation. This calibration separates the existence of a static shell from the subsequent test of its local radial stability and avoids scanning parameter combinations that do not reproduce the junction stresses. Representative parameter families are used to display how the string strength, regularization scale, throat position, and fluid parameters modify the surface stresses and the sign of the potential curvature. We also use the exact unregularized limit as an analytic control to determine which qualitative shell properties persist when the rational regulator is removed. Physically, the construction isolates the additional exotic source on the throat while preserving the regular string-cloud geometry in the bulk, allowing the roles of regularization, asymptotic string matter, and shell microphysics to be compared within a single framework. As a complementary analysis, we use backward ray tracing to examine celestial-sphere lensing and thin-disk images, varying the string-cloud strength, throat radius, and viewing inclination. We also compare one-sided and two-sided disk models to distinguish higher-order images of the observer-side disk from genuine through-throat images of matter in the second exterior.

This paper is organized as follows. Section~\ref{sec:seed} reviews the regularized Letelier--Alencar black-hole geometry, its core and asymptotic limits, and the admissible positive-lapse exterior. Section~\ref{sec:construction} constructs the symmetric junction and derives the induced geometry, unit normals, and extrinsic curvature. Section~\ref{sec:surface} obtains the surface stresses and analyzes the energy conditions. Section~\ref{sec:stability} develops the conservation law, effective potential, and general local-stability criterion. Section~\ref{sec:eos} applies the formalism to the five surface equations of state and discusses representative stability profiles. Section~\ref{sec:optical_appearance} introduces the thin-disk model and backward-ray-tracing method and analyzes the lensing and radiative images of the wormhole. Section~\ref{sec:conclusion} summarizes the results and limitations. We use geometrized units $G=c=1$.

\section{Regularized Letelier--Alencar string-cloud black hole}\label{sec:seed}

The seed geometry is the exact static and spherically symmetric solution introduced in Ref.~\cite{Muniz2026}, obtained by applying a rational Dagum-type regulator to the Letelier--Alencar string-cloud lapse of Ref.~\cite{Alencar2025}. We denote the non-negative, dimensionless string-cloud strength by $\mathcal A$. The line element is written directly as
\begin{equation}
\dd s^2=-F(r)\dd t^2+\frac{\dd r^2}{F(r)}+r^2\left(\dd\theta^2+\sin^2\theta\,\dd\phi^2\right),
\label{eq:seed_metric}
\end{equation}
with
\begin{equation}
\begin{aligned}
F(r)=1
-&\left[
\frac{2M}{r}
-\frac{\mathcal A r_0^2}{r^2}
\Fh\!\left(-\frac12,-\frac14;\frac34;-\frac{r^4}{r_0^4}\right)
\right]
\left(1+\frac{r_0}{r}\right)^{-4}.
\end{aligned}
\label{eq:seed_lapse}
\end{equation}
Here $r$ is the areal radius, because the area of each orbit of spherical symmetry is $4\pi r^2$. The coordinate $t$ is the static time coordinate, $(\theta,\phi)$ are the standard angular coordinates, and ${}_2F_1$ denotes the Gaussian hypergeometric function. The positive mass parameter $M>0$ and the regularization scale $r_0>0$ have dimensions of length, whereas $\mathcal A$ is dimensionless. The parameter $r_0$ controls the size of the region over which the central source is smeared. The string parameter $\mathcal A$ controls both the curvature scale of the regular core and the leading string-cloud contribution at large radius. The mass $M$ fixes the central Schwarzschild-like attraction but does not by itself determine the horizon structure, because the roots also depend on $\mathcal A$ and $r_0$.

The central behavior follows from the expansion
\begin{equation}
F(r)=1+\frac{\mathcal A}{r_0^2}r^2+\mathcal O(r^3),
\qquad r\rightarrow0.
\end{equation}
For $\mathcal A>0$, the core is locally anti--de Sitter, with curvature radius and effective cosmological constant, respectively, $r_0/\sqrt{\mathcal A}$ and $-3\mathcal A/r_0^2$. In particular, the Kretschmann scalar approaches $24\mathcal A^2/r_0^4$, rather than diverging. The metric is therefore regular at $r=0$. This property is crucial for the interpretation of the seed solution, although the thin-shell construction developed below excises the entire region below the throat and hence does not rely on passage through the core.

At large radius, the lapse has the leading form

\begin{equation}
F(r)=1-\mathcal A-\frac{2\widetilde M}{r}
+\mathcal O(r^{-2}),
\end{equation}

where
\begin{equation}
\widetilde M
=M-\frac{r_0\Gamma(1/4)\mathcal A}{2\Gamma(5/4)}
+\frac{r_0\Gamma(-1/4)\Gamma(3/4)\mathcal A}{4\sqrt{\pi}}.
\end{equation}
Here $\widetilde M$ is the effective mass coefficient of the asymptotic $1/r$ term, has dimensions of length, and $\Gamma$ denotes Euler's gamma function.
Thus the regularized geometry approaches the Letelier-type string-cloud exterior rather than the Minkowski metric whenever $\mathcal A\neq0$ \cite{Letelier1979,Muniz2026}. The constant limit $F(\infty)=1-\mathcal A$ can be absorbed from the temporal and radial sectors by constant rescalings, but the areal sector then retains a solid-angle defect. Consequently, the term ``asymptotically flat'' must be understood in the looser string-cloud sense used in parts of the literature; strict asymptotic flatness and a standard ADM normalization are recovered only for $\mathcal A=0$. For the two-ended static exterior used here, we impose
\begin{equation}
0\leq \mathcal A<1,
\end{equation}
so that the lapse remains positive sufficiently far from the compact object.

The horizons are the positive roots of
\begin{equation}
F(r_h)=0.
\end{equation}
Because Eq.~\eqref{eq:seed_lapse} contains a Gaussian hypergeometric function, the roots must in general be found numerically. Depending on $(M,\mathcal A,r_0)$, the geometry may possess two positive roots, one degenerate root, or no root \cite{Muniz2026}. In the black-hole sector studied in this work, we denote the larger root by $r_+$ and retain the connected exterior region
\begin{equation}
r>r_+,
\qquad F(r)>0.
\label{eq:retained_domain}
\end{equation}
A degenerate horizon lies on the boundary of the black-hole parameter domain and satisfies $F(r_e)=F'(r_e)=0$. The throat will always be placed strictly inside the open static region, not on a simple or degenerate horizon. The metric then has Lorentzian signature $(-,+,+,+)$ and the shell can be timelike.

Three limits separate the respective roles of the regulator and the string cloud. First, consider $r_0\rightarrow0$ at fixed $M$ and $\mathcal A$. We denote quantities evaluated in this unregularized exterior by the roman label $\mathrm{u}$; this label is a mnemonic and not a tensor index. In this limit, Eq.~\eqref{eq:seed_lapse} reduces to

\begin{equation}
F_{\mathrm{u}}(r)=1-\mathcal A-\frac{2M}{r},
\label{eq:unregularized_lapse}
\end{equation}

which is the lapse of the singular Letelier string-cloud exterior. For $M>0$ and $0\leq\mathcal A<1$, this lapse has a single positive horizon. Let $a_0$ denote the equilibrium areal radius of a prospective static shell. The horizon radius and the corresponding dimensionless throat position are
\begin{equation}
r_+^{\mathrm{u}}=\frac{2M}{1-\mathcal A},
\qquad
x_{\mathrm{u}}=\frac{a_0}{r_+^{\mathrm{u}}}>1,
\label{eq:unregularized_scales}
\end{equation}
 respectively. Thus, $x_{\mathrm{u}}>1$ is precisely the retained-exterior condition $a_0>r_+^{\mathrm{u}}$; the shell geometry and its static limit are developed in Sec.~\ref{sec:construction}.

Second, taking $\mathcal A\rightarrow0$ after this limit gives the Schwarzschild solution. Third, setting $\mathcal A=0$ at finite $r_0$ leaves a rationally regularized Schwarzschild-type geometry rather than the exact Schwarzschild lapse. These limits should not be conflated. They show separately how the string cloud and the regulator modify the exterior and the core.

Only one property of the bulk stress tensor in the Schwarzschild-like gauge of Eq.~\eqref{eq:seed_metric} is required below.
Writing the effective bulk energy--momentum tensor as $T^\mu{}_{\nu}$, with energy density $\rho\equiv-T^t{}_{t}$ and radial pressure $p_r\equiv T^r{}_{r}$, the equality $T^t{}_{t}=T^r{}_{r}$ implies the radial-vacuum relation $p_r=-\rho$ \cite{Muniz2026}. This radial vacuum relation makes the net momentum flux across the symmetric moving shell vanish. No other bulk stress components are required for the junction analysis.

In addition, as noted previously, the distinction between exterior and core effects is important for interpreting the results below. The thin-shell construction retains only the region outside the throat and excises both the horizons and the regular core. Consequently, the energy-condition and radial-stability results probe the rational regularization only through its modification of the retained exterior lapse; they do not show that the nonsingular core itself improves either property. We therefore use the exact $r_0\rightarrow0$ exterior as an analytic control in the discussions below. This fixed-$(M,\mathcal A)$ limit tests whether the qualitative shell behavior survives without regularization, but it is not by itself a mass-matched isolation of the regulator because both the asymptotic mass coefficient and the horizon normalization change with $r_0$.

\section{Symmetric excision and throat geometry}\label{sec:construction}

We construct the wormhole from two identical copies of the exterior region in Eq.~\eqref{eq:retained_domain}. The shell history is parametrized by its proper time $\tau$, and $a(\tau)$ denotes its dynamical areal radius. Throughout the junction and stability analysis, an overdot denotes $\dd/\dd\tau$, while a prime on the lapse denotes $\dd/\dd r$. On each copy, the portion below a spherical timelike boundary is removed, and the two remaining boundaries are identified. The retained manifolds are
\begin{equation}
\mathcal M^{\pm}=\left\{x_\pm^\mu=(t_\pm,r_\pm,\theta,\phi)\,\middle|\,
 r_\pm\geq a(\tau)>r_+,\ F[a(\tau)]>0\right\}.
\end{equation}
Their common boundary is the shell $\Sigma:r=a(\tau)$. The intrinsic coordinates are $\xi^i=(\tau,\theta,\phi)$, and the embeddings are
\begin{equation}
x_\pm^\mu(\tau,\theta,\phi)=\bigl(t_\pm(\tau),a(\tau),\theta,\phi\bigr).
\end{equation}
The proper-time normalization of the shell four-velocity is explicitly
\begin{equation}
-F(a)\dot t_\pm^{\,2}+\frac{\dot a^{\,2}}{F(a)}=-1,
\qquad
\dot t_\pm=\frac{\sqrt{F(a)+\dot a^2}}{F(a)},
\label{eq:proper_time_relation}
\end{equation}
where the future-directed root has been selected. Since $F(a)>0$, the induced temporal direction remains timelike for every finite $\dot a$ allowed by the local dynamics.

The first fundamental form on $\Sigma$ is the metric induced by the pullback of the bulk geometry. On each side of the junction it is given by
\begin{equation}
h^{\pm}_{ij}
=
g^{\pm}_{\mu\nu} e^{\mu}_{i,\pm} e^{\nu}_{j,\pm},
\qquad
e^{\mu}_{i,\pm}
=
\frac{\partial x^\mu_\pm}{\partial \xi^i}.
\end{equation}
Here $g^{\pm}_{\mu\nu}$ denotes the bulk metric on the corresponding copy $\mathcal M^\pm$, evaluated on the shell, while $e^{\mu}_{i,\pm}$ are the tangent vectors to $\Sigma$ induced by the embedding $x^\mu_\pm(\xi^i)$. Greek indices refer to spacetime coordinates, whereas Latin indices label coordinates intrinsic to the shell. Using the embeddings introduced above, together with the proper-time condition for the shell trajectory, the nonvanishing components of the induced metric are
\begin{equation}
h^{\pm}_{\tau\tau}
=
-F(a)\dot t_\pm^{\,2}
+
\frac{\dot a^{\,2}}{F(a)}
=
-1,
\qquad
h^{\pm}_{\theta\theta}
=
a^2(\tau),
\qquad
h^{\pm}_{\phi\phi}
=
a^2(\tau)\sin^2\theta .
\end{equation}
Therefore, the intrinsic line element of the shell is
\begin{equation}
\dd s_{\Sigma}^{2}
=
h_{ij}\dd\xi^i\dd\xi^j
=
-\dd\tau^2
+
a^2(\tau)
\left(
\dd\theta^2+\sin^2\theta\,\dd\phi^2
\right).
\end{equation}
The first junction condition,
\begin{equation}
[h_{ij}]
\equiv
h^+_{ij}-h^-_{ij}
=
0,
\end{equation}
is satisfied identically in the present construction. Indeed, both sides are copies of the same exterior geometry and are identified at the same areal radius $r=a(\tau)$.

On either bulk chart, a local level-set representation compatible with the embedding is $f_\pm(t_\pm,r_\pm)=r_\pm-a[\tau(t_\pm)]$, where the proper-time relation makes $\tau(t_\pm)$ locally well defined along the shell; below, $f$ denotes the corresponding function on either side. On the shell, $\partial_{t_\pm}f_\pm=-\dot a/\dot t_\pm$ and $\partial_{r_\pm}f_\pm=1$, so Eq.~\eqref{eq:proper_time_relation} gives the explicit normal below. The outward-oriented unit normal covectors are obtained from its normalized gradient and, for the present metric, take the explicit form
\begin{equation}
\begin{aligned}
n_\mu^{\pm}
&=\pm\left|g^{\alpha\beta}\partial_\alpha f\,\partial_\beta f\right|^{-1/2}\partial_\mu f
=\pm\left(-\dot a,\frac{\sqrt{F(a)+\dot a^2}}{F(a)},0,0\right).
\end{aligned}
\label{eq:unit_normal}
\end{equation}
The sign distinguishes the outward direction on the two copies. The normals are orthogonal to the shell trajectory and have unit spacelike norm. These properties follow directly from the proper-time relation used above and from the normalization in Eq.~\eqref{eq:unit_normal}.

The second fundamental form is identified as the extrinsic curvature, given by
\begin{equation}
K_{ij}^{\pm}=e_i^{\mu}e_j^{\nu}\Del_\mu n_\nu^{\pm}.
\end{equation}
In an orthonormal basis on the shell, its independent components are
\begin{equation}
K^{\pm\,\hat\tau}{}_{\hat\tau}
=\pm\frac{F'(a)+2\ddot a}{2\sqrt{F(a)+\dot a^2}},
\qquad
K^{\pm\,\hat\theta}{}_{\hat\theta}
=K^{\pm\,\hat\phi}{}_{\hat\phi}
=\pm\frac{\sqrt{F(a)+\dot a^2}}{a}.
\label{eq:extrinsic_components}
\end{equation}
For a static throat, $a=a_0$ and $\dot a=\ddot a=0$, so that
\begin{equation}
\left.K^{\pm\,\hat\tau}{}_{\hat\tau}\right|_0
=\pm\frac{F'_0}{2\sqrt{F_0}},
\qquad
\left.K^{\pm\,\hat\theta}{}_{\hat\theta}\right|_0
=\left.K^{\pm\,\hat\phi}{}_{\hat\phi}\right|_0
=\pm\frac{\sqrt{F_0}}{a_0},
\end{equation}
where $F_0=F(a_0)$ and $F'_0=F'(a_0)$. More generally, a subscript $0$ denotes evaluation at the static throat $a=a_0$, while a subscript $+$ denotes evaluation at the outer horizon $r=r_+$.

Geometrically, the cut removes $r<a(\tau)$ from both copies, including the horizons and regular core whenever the shell remains outside $r_+$. The areal radius increases away from the junction on both sides. Hence the sphere at $r=a$ is a minimum-area surface in the glued manifold and realizes the radial flare-out of the thin-shell throat. The construction does not by itself establish global geodesic completeness or stability under nonspherical perturbations; those stronger properties require separate analyses.

\section{Surface stresses and energy conditions}\label{sec:surface}

The discontinuity of the normal derivative of the metric is represented by a distributional surface source. In a comoving orthonormal basis, spherical symmetry gives
\begin{equation}
S^{\hat i}{}_{\hat j}=\operatorname{diag}(-\sigma,p,p),
\label{eq:surface_tensor_orthonormal}
\end{equation}
where $\sigma$ is the surface energy density and $p$ is the tangential pressure.

With the normal convention of Eq.~\eqref{eq:unit_normal}, the Lanczos equation is \cite{Israel1966,Visser1989}
\begin{equation}
S^{\hat i}{}_{\hat j}
=-\frac{1}{8\pi}
\left([K^{\hat i}{}_{\hat j}]
-\delta^{\hat i}{}_{\hat j}[K]\right),
\qquad
[K^{\hat i}{}_{\hat j}]
=K^{+\,\hat i}{}_{\hat j}-K^{-\,\hat i}{}_{\hat j}.
\end{equation}
Here $[K]=[K^{\hat k}{}_{\hat k}]$ is the trace of the jump evaluated in the same orthonormal basis.
Substitution of Eq.~\eqref{eq:extrinsic_components} yields
\begin{equation}
\sigma(a,\dot a)=-\frac{1}{2\pi a}\sqrt{F(a)+\dot a^2},
\label{eq:dynamic_sigma}
\end{equation}
\begin{equation}
p(a,\dot a,\ddot a)=\frac{1}{8\pi}
\left[
\frac{F'(a)+2\ddot a}{\sqrt{F(a)+\dot a^2}}
+\frac{2\sqrt{F(a)+\dot a^2}}{a}
\right].
\end{equation}
The static values are
\begin{equation}
\sigma_0=-\frac{\sqrt{F_0}}{2\pi a_0},
\qquad
p_0=\frac{a_0F'_0+2F_0}{8\pi a_0\sqrt{F_0}}.
\label{eq:static_stresses}
\end{equation}
Since the throat lies in the positive-lapse exterior, $F_0>0$ and therefore $\sigma_0<0$. The sign is not inherited from the regular string-cloud matter in the bulk. It is generated by the jump of the angular extrinsic curvature required to make the areal radius increase away from the shell on both sides. The exotic matter supporting the wormhole is thus localized on $\Sigma$, whereas the bulk on each side remains the original regular black-hole solution.

\begin{figure*}[htp!]
    \centering
    \includegraphics[width=0.98\linewidth]{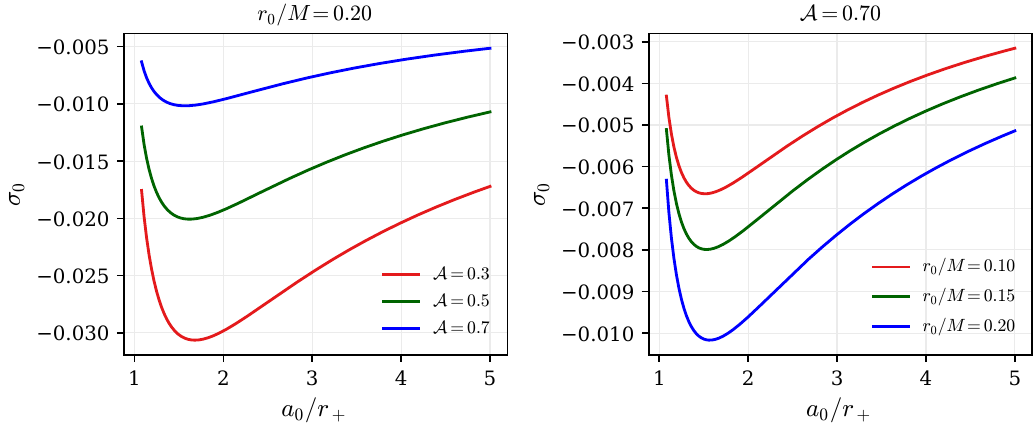}
    \caption{Static surface energy density $\sigma_0$ as a function of the dimensionless throat radius $a_0/r_+$. In the left panel, $r_0/M=0.20$ is fixed and the curves correspond to $\mathcal A=0.3,0.5,0.7$. In the right panel, $\mathcal A=0.70$ is fixed and the curves correspond to $r_0/M=0.10,0.15,0.20$. In both panels, $M=1$. The curves are shown only in the admissible positive-lapse exterior $a_0>r_+$, where the static thin-shell construction is well defined.}
    \label{fig:static_surface_density}
\end{figure*}

\begin{figure*}[htp!]
    \centering
    \includegraphics[width=0.98\linewidth]{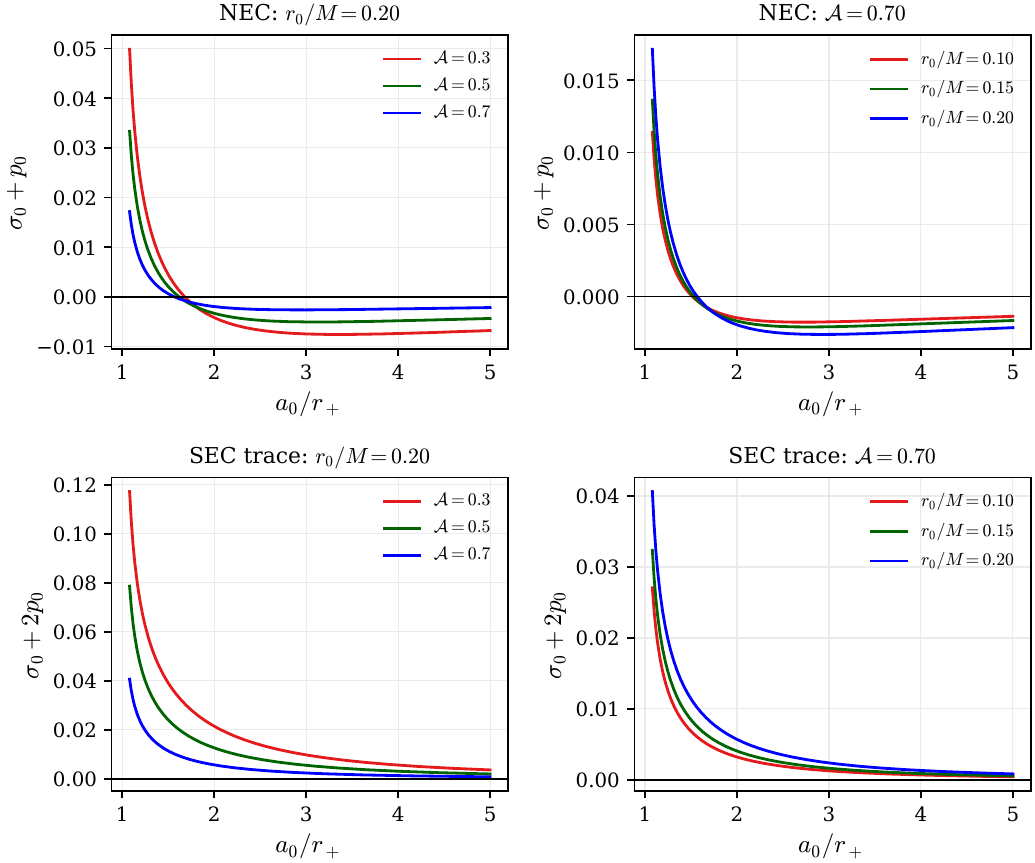}
    \caption{Static energy-condition combinations as functions of the dimensionless throat radius $a_0/r_+$. The upper panels show the tangential NEC combination $\sigma_0+p_0$, while the lower panels show the auxiliary density-plus-tangential-pressures combination $\sigma_0+2p_0$. In the left column, $r_0/M=0.20$ is fixed and the curves correspond to $\mathcal A=0.3,0.5,0.7$; in the right column, $\mathcal A=0.70$ is fixed and the curves correspond to $r_0/M=0.10,0.15,0.20$. In all panels, $M=1$ and only the admissible positive-lapse exterior is retained. The horizontal line marks the zero level separating satisfied and violated combinations.}
    \label{fig:static_energy_conditions}
\end{figure*}

The profiles in Fig.~\ref{fig:static_surface_density} confirm the sign fixed analytically by Eq.~\eqref{eq:static_stresses}: every admissible static shell has negative surface energy density. Each curve emerges from the outer-horizon limit with $\sigma_0\rightarrow0^{-}$, reaches a single negative minimum, and then rises gradually toward zero. The first behavior follows from $F_0\rightarrow0^{+}$, whereas the outer trend is consistent with the finite string-cloud limit of the lapse and the explicit inverse-radius factor in Eq.~\eqref{eq:static_stresses}. Differentiating that equation also shows that an extremum of $\sigma_0$ is controlled by $a_0F'_0-2F_0$; consequently, the minimum occurs at the same radius at which the null energy condition changes sign in Fig.~\ref{fig:static_energy_conditions}.

At fixed $r_0/M$, increasing $\mathcal A$ shifts the curves upward, reduces the magnitude of the negative minimum, and moves it modestly toward smaller $a_0/r_+$. The opposite ordering is produced by the regularization scale in the right panel: increasing $r_0/M$ deepens the minimum and shifts it toward larger normalized radii. Since the horizontal coordinate is normalized by the parameter-dependent outer horizon, these trends combine changes in the local lapse with substantial changes in $r_+$; they should therefore not be attributed to either factor separately. The sign of the shell density is unaffected, but the magnitude of the negative surface contribution is sensitive to both parameters in the range displayed.

The unregularized control makes the persistence of this qualitative profile explicit without requiring an additional plotted curve. Substitution of Eq.~\eqref{eq:unregularized_lapse} into Eq.~\eqref{eq:static_stresses} gives
\begin{equation}
\sigma_{0,\mathrm{u}}
=-\frac{(1-\mathcal A)^{3/2}}{4\pi M}
\frac{\sqrt{x_{\mathrm{u}}-1}}{x_{\mathrm{u}}^{3/2}}<0,
\qquad x_{\mathrm{u}}>1.
\label{eq:unregularized_surface_density}
\end{equation}
This expression tends to zero at both the outer-horizon boundary and large radius and has a single intervening minimum. The regularized profiles therefore connect to the same sign and radial topology; regularization changes the scale and location of the negative surface contribution, but it does not remove the exotic density imposed by the symmetric junction.

This reduction also gives a direct formula-level comparison with the singular Letelier thin shell of Richarte and Simeone \cite{RicharteSimeone2008}. Under the notation map from their string parameter and throat radius to $(\mathcal A,a_0)$, Eq.~\eqref{eq:unregularized_surface_density} coincides with their static surface-density law. It also reproduces their simple-horizon limit, in which the negative density tends to zero while the tangential pressure diverges. The agreement provides an external algebraic consistency check of the junction sign and the exact $r_0=0$ reduction; the finite-$r_0$ displacement of the profiles remains a result of the present regularized lapse.

For a static surface layer in $2+1$ dimensions, the relevant combinations are
\begin{equation}
\sigma_0+p_0
=\frac{a_0F'_0-2F_0}{8\pi a_0\sqrt{F_0}},
\qquad
\sigma_0+2p_0
=\frac{F'_0}{4\pi\sqrt{F_0}}.
\end{equation}

Here intrinsic energy conditions are evaluated using causal vectors tangent to the $2+1$-dimensional shell. The intrinsic null energy condition (NEC) requires $\sigma_0+p_0\geq0$. The intrinsic weak energy condition (WEC) additionally requires $\sigma_0\geq0$, and the intrinsic dominant energy condition (DEC) requires non-negative density together with $\sigma_0\geq|p_0|$; both are therefore violated for every admissible static throat. The standard dimension-dependent definition of the intrinsic strong energy condition (SEC), $(S_{ij}-S h_{ij})v^iv^j\geq0$ for every timelike tangent vector $v^i$, where $S=S^i{}_i=-\sigma+2p$, reduces in $2+1$ dimensions to $p_0\geq0$ together with $\sigma_0+p_0\geq0$. If instead the surface tensor is embedded in four dimensions with zero normal pressure, the density-plus-principal-pressures combination is $\sigma_0+2p_0$; this motivates the auxiliary quantity displayed above but does not replace the complete four-dimensional energy-condition tests. In the present symmetric construction, $\sigma_0<0$, so $\sigma_0+p_0\geq0$ already implies $p_0>0$ and $\sigma_0+2p_0>0$. The standard intrinsic SEC consequently has the same radial domain as the tangential NEC. The distinction remains physical: a four-dimensional null direction with a nonzero normal component probes a contribution proportional to the negative $\sigma_0$, so the complete distributional NEC is violated even where its tangential restriction is satisfied.

Approaching a simple outer horizon from the exterior illustrates why $\sigma_0\rightarrow0$ does not imply the disappearance of the shell. If $F(r_+)=0$ and $F'(r_+)>0$, then $\sigma_0$ tends to zero from below, while the leading term in $p_0$ is proportional to $F'_+ /\sqrt{F_0}$ and diverges. The surface layer therefore becomes pressure dominated. A degenerate horizon must be treated separately because $F'_+=0$ changes the limiting balance. In either case, the throat is kept strictly outside the root so that the static junction remains timelike and regular.

The upper panels of Fig.~\ref{fig:static_energy_conditions} separate two radial regimes. Immediately outside the outer horizon, $\sigma_0+p_0$ is positive and grows rapidly as the lower boundary is approached. It then crosses zero once and remains negative over the outer part of the displayed exterior, approaching zero from below. The near-horizon growth is pressure driven: for a simple root, the terms proportional to $F'_0/\sqrt{F_0}$ become large even though $\sigma_0\rightarrow0^{-}$. Thus, the positive tangential NEC combination close to $r_+$ does not represent a regular matter-free limit. Its zero coincides with the minimum of $\sigma_0$ in Fig.~\ref{fig:static_surface_density}, as follows from the common factor $a_0F'_0-2F_0$.

The auxiliary combination $\sigma_0+2p_0$ remains positive throughout both lower panels and decreases toward zero as the throat is moved outward, indicating that $F'_0$ stays positive over the sampled exterior. Equations~\eqref{eq:static_stresses} then also give $p_0>0$ because $F_0>0$. The standard intrinsic SEC is therefore satisfied precisely in the inner portion where the tangential NEC holds. This statement is confined to the shell tangent space; the complete four-dimensional distributional NEC remains violated because $\sigma_0<0$. The intrinsic WEC and DEC likewise remain violated everywhere. Increasing $\mathcal A$ reduces the magnitudes of both combinations and shifts the tangential NEC transition slightly inward in $a_0/r_+$, whereas increasing $r_0/M$ enhances the near-horizon amplitudes and moves that transition modestly outward. These statements apply to the parameter families and radial interval shown.

The same division into inner and outer regimes already occurs in the singular string-cloud exterior. Indeed,
\begin{equation}
a_0F'_{\mathrm{u}}(a_0)-2F_{\mathrm{u}}(a_0)=0
\quad\Longleftrightarrow\quad
a_0=\frac{3M}{1-\mathcal A}=\frac{3}{2}r_+^{\mathrm{u}},
\label{eq:unregularized_nec_transition}
\end{equation}
whereas $F'_{\mathrm{u}}>0$ throughout the exterior. Hence the unregularized shell has a positive tangential NEC combination only inside that transition and a positive auxiliary combination $\sigma_{0,\mathrm{u}}+2p_{0,\mathrm{u}}$ everywhere, while its negative density continues to exclude the intrinsic WEC and DEC and the complete distributional NEC. Because $p_{0,\mathrm{u}}>0$, the intrinsic SEC has the same inner domain as the tangential NEC. The regularized curves are therefore consistent with a continuous deformation of the same intrinsic energy-condition classification rather than evidence that regularity of the excised core creates a new admissible regime.

\section{Conservation, radial dynamics, and local perturbations}\label{sec:stability}

Because the Lanczos equation is tensorial, $S^i{}_{j}$ below denotes the components of the same surface tensor in the intrinsic coordinate basis; these are related to the hatted orthonormal components in Eq.~\eqref{eq:surface_tensor_orthonormal} by the shell vielbein. The contracted Codazzi equation gives the intrinsic conservation law \cite{Israel1966,PoissonVisser1995}
\begin{equation}
\Del_i S^i{}_{j}=-\left[T_{\mu\nu}e^\mu{}_j n^\nu\right].
\label{eq:surface_conservation}
\end{equation}
Using the bulk definitions $\rho=-T^t{}_{t}$ and $p_r=T^r{}_{r}$ given in Sec.~\ref{sec:seed}, the radial-vacuum identity $T^t{}_{t}=T^r{}_{r}$ implies $\rho+p_r=0$. The normal energy flux is proportional to this combination and therefore vanishes on each copy; its jump consequently vanishes for the symmetric junction. The temporal component of Eq.~\eqref{eq:surface_conservation} reduces to
\begin{equation}
\dot\sigma+\frac{2\dot a}{a}(\sigma+p)=0,
\qquad
\sigma'(a)=-\frac{2}{a}(\sigma+p),
\label{eq:radial_conservation}
\end{equation}
Here the prime on $\sigma$ denotes differentiation along a solution of the radial conservation equation with the equation-of-state parameters held fixed. It must not be interpreted as differentiation along the sequence of separately recalibrated static equilibria used in the figures.

Solving Eq.~\eqref{eq:dynamic_sigma} for $\dot a^2$ gives the one-dimensional radial equation of motion, in which $V(a)$ is the effective potential governing the shell radius,
\begin{equation}
\dot a^2+V(a)=0,
\qquad
V(a)=F(a)-4\pi^2a^2\sigma^2(a).
\label{eq:effective_potential}
\end{equation}
For a general surface equation of state $p=p(\sigma,a)$, we define
\begin{equation}
\zeta(a)=\left(\frac{\partial p}{\partial\sigma}\right)_a,
\qquad
p_a(a)=\left(\frac{\partial p}{\partial a}\right)_\sigma.
\end{equation}
Two differentiations of Eq.~\eqref{eq:effective_potential}, together with Eq.~\eqref{eq:radial_conservation}, give
\begin{equation}
\begin{aligned}
V''(a)=F''(a)+8\pi^2\bigl\{&\sigma(\sigma+2p)
-4(\sigma+p)\left[\sigma+p+\sigma\zeta\right]
+2a\sigma p_a\bigr\}.
\end{aligned}
\end{equation}
At a static radius, the junction values in Eq.~\eqref{eq:static_stresses} reduce this expression to
\begin{equation}
V''(a_0)=F''_0-\frac{F'^2_0}{2F_0}
+\left(1+2\zeta_0\right)
\left(\frac{F'_0}{a_0}-\frac{2F_0}{a_0^2}\right)
-8\pi\sqrt{F_0}\,p_{a0}.
\label{eq:static_vpp_general}
\end{equation}
Here $\zeta_0\equiv\zeta(a_0)$ and $p_{a0}\equiv p_a(a_0)$ are evaluated for the equation of state calibrated at that equilibrium.
No expanded formula for $F'$ or $F''$ is needed: they are obtained by differentiating the exact lapse in Eq.~\eqref{eq:seed_lapse} symbolically or numerically at the selected radius.

A static solution must satisfy $V(a_0)=0$ and $V'(a_0)=0$. The first condition follows directly from $\sigma_0$ in Eq.~\eqref{eq:static_stresses}; the second follows when the chosen equation of state reproduces $p_0$ at the same radius. We perturb the shell according to
\begin{equation}
a(\tau)=a_0+\delta a(\tau),
\qquad |\delta a|\ll a_0,
\end{equation}
and expand the potential to second order:
\begin{equation}
V(a)=V(a_0)+V'(a_0)\delta a
+\frac12V''(a_0)(\delta a)^2+\mathcal O((\delta a)^3).
\end{equation}
Differentiating the equation of motion and retaining linear terms yields
\begin{equation}
\ddot{\delta a}+\frac12V''(a_0)\delta a=0.
\end{equation}
Thus $V''(a_0)>0$ gives local oscillatory stability, $V''(a_0)<0$ gives local exponential instability, and $V''(a_0)=0$ requires higher-order analysis. A negative second derivative does not by itself prove global collapse. Moreover, changing $a_0$ in a parameter scan selects a different calibrated static wormhole; it is not the same operation as evolving the perturbation $\delta a(\tau)$ around one fixed equilibrium.

\section{Surface equations of state}\label{sec:eos}

The following equations of state are phenomenological closures for the distributional matter on the shell. For each model, the static junction condition is used to determine one natural equation-of-state parameter. Once calibrated at a chosen $a_0$, that parameter is held fixed while the corresponding wormhole is perturbed.

\subsection{Linear barotropic fluid}\label{subsec:barotropic}

The linear closure
\begin{equation}
p=w_{\rm B}\sigma
\label{eq:barotropic_eos}
\end{equation}
is widely used in thin-shell stability analyses \cite{PoissonVisser1995,Kuhfittig2010,Javed2024}. The coefficient $w_{\rm B}$ is dimensionless. In cosmology, particular negative values are often called dark-energy or phantom parameters, but that interpretation is not automatic here because $\sigma_0<0$ and the matter is a surface layer. Its directly verifiable role is to fix the local pressure response to a change in surface density.

For Eq.~\eqref{eq:barotropic_eos}, $\zeta=w_{\rm B}$ and $p_a=0$. The static condition $p_0=w_{\rm B}\sigma_0$ fixes
\begin{equation}
w_{\rm B}=-\frac{a_0F'_0+2F_0}{4F_0}.
\label{eq:barotropic_calibration}
\end{equation}
Substitution into Eq.~\eqref{eq:static_vpp_general} gives
\begin{equation}
V''_{\rm B}(a_0)=F''_0+\frac{F'_0}{a_0}-\frac{F'^2_0}{F_0}.
\label{eq:barotropic_vpp}
\end{equation}

\begin{figure*}[htp!]
    \centering
    \includegraphics[width=0.98\linewidth]{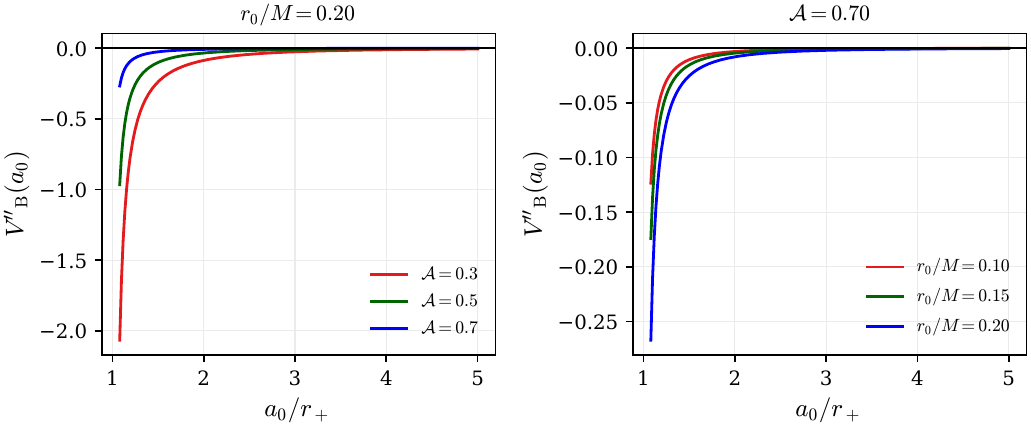}
    \caption{Linear barotropic-fluid stability function $V''_{\rm B}(a_0)$ as a function of $a_0/r_+$. In the left panel, $r_0/M=0.20$ is fixed and the curves correspond to $\mathcal A=0.3,0.5,0.7$; in the right panel, $\mathcal A=0.70$ is fixed and the curves correspond to $r_0/M=0.10,0.15,0.20$. In both panels, $M=1$, and $w_{\rm B}$ is fixed by the static junction condition at each displayed throat. The curves are restricted to the positive-lapse exterior. The horizontal zero level separates locally stable configurations, $V''>0$, from unstable ones, $V''<0$.}
    \label{fig:stability_barotropic}
\end{figure*}

No barotropic branch in Fig.~\ref{fig:stability_barotropic} crosses the zero level. The potential curvature is strongly negative close to the outer horizon and then increases monotonically toward zero from below, so the displayed equilibria are locally unstable throughout the sampled exterior. The sharp inner behavior follows directly from Eq.~\eqref{eq:barotropic_vpp}: near a simple horizon, the term $-F_0'^2/F_0$ dominates and drives $V''_{\rm B}$ to negative values. Farther from the compact object, the remaining geometrical terms decay and the curves approach marginality without developing a positive branch.

At fixed $r_0/M$, larger $\mathcal A$ substantially reduces the magnitude of the negative curvature. For fixed $\mathcal A$, increasing the regularization scale has the opposite effect close to the horizon, although the three curves become nearly indistinguishable at larger $a_0/r_+$. These shifts change the severity of the local instability but not its sign in the plotted range. Moreover, each point corresponds to a different static shell with $w_{\rm B}$ recalibrated through Eq.~\eqref{eq:barotropic_calibration}; the curves are therefore families of equilibria rather than temporal trajectories of one wormhole.

This absence of a stable branch is not produced by the rational regulator. In the unregularized exterior, Eq.~\eqref{eq:barotropic_vpp} reduces exactly to
\begin{equation}
V''_{\mathrm{B,u}}(a_0)
=-\frac{(1-\mathcal A)^3}
{4M^2x_{\mathrm{u}}^2(x_{\mathrm{u}}-1)}<0,
\qquad x_{\mathrm{u}}>1.
\label{eq:barotropic_vpp_unregularized}
\end{equation}
The finite-$r_0$ families preserve this analytic sign structure and mainly alter the strength of the instability, supporting a smooth limiting connection rather than a regulator-induced change of stability class.

The broader stability analysis of the same singular Letelier exterior by Richarte and Simeone \cite{RicharteSimeone2008} is consistent with this result once the equation-of-state restrictions are matched. Setting $r_0=0$ and $p_{a0}=0$ in Eq.~\eqref{eq:static_vpp_general} reproduces their criterion after identifying their free local slope $\eta_0=(\partial p/\partial\sigma)_0$ with $\zeta_0$. The linear closure used here imposes the additional calibration $\eta_0=w_{\rm B}=p_0/\sigma_0$, selecting a definite slice of their more general stability plane; on that slice, Eq.~\eqref{eq:barotropic_vpp_unregularized} proves $V''_{\rm B}<0$ throughout the exterior. The parameter-dependent barotropic stable sectors reported for Bardeen--AdS and Hayward--Letelier--AdS string-fluid seeds \cite{WaseemEtAl2025StringFluid,HouedokouEtAl2026HaywardLetelier} therefore do not contradict the present result: their charge, anti--de Sitter, and regularization terms change the lapse derivatives entering the same local criterion.

Equation~\eqref{eq:barotropic_calibration} shows that one cannot vary $a_0$ while keeping the same calibrated $w_{\rm B}$ unless the new radius independently satisfies the same static relation. Each generic radius therefore represents a distinct barotropic shell.

\subsection{Variable phantom equation of state}\label{subsec:phantom}

Varela introduced an explicit radius dependence into the linear shell law in order to enlarge the possible stability behavior of Schwarzschild thin shells \cite{Varela2015}. We write the model as
\begin{equation}
p=\frac{\Delta_{\rm P}}{a^n}\sigma,
\end{equation}
Here $\Delta_{\rm P}\in\mathbb R$ is the amplitude of the radius-dependent pressure response, with $[\Delta_{\rm P}]=L^n$, and $n$ is a dimensionless real exponent. The figures sample $1.5\leq n\leq3$; $\Delta_{\rm P}$ is not scanned independently, because its value and sign are recalibrated at every displayed static throat. The case $n=0$ reduces to the constant linear law. Mathematically, $n$ may be real; the original Schwarzschild analysis emphasized $n>0$ because that range produced a nontrivial stability boundary. In the present geometry the sign of $V''$ must be reassessed from the actual lapse, so no stability interval is imported from the Schwarzschild calculation.

The required derivatives are
\begin{equation}
\zeta_{\rm P}=\frac{\Delta_{\rm P}}{a^n}=\frac{p}{\sigma},
\qquad
p_{a}^{\rm(P)}=-\frac{n}{a}p.
\end{equation}
The static condition fixes the amplitude rather than the exponent,
\begin{equation}
\Delta_{\rm P}
=a_0^n\frac{p_0}{\sigma_0}
=-\frac{a_0^n(a_0F'_0+2F_0)}{4F_0}.
\label{eq:phantom_calibration}
\end{equation}
This is the natural choice because $n$ controls the prescribed radial dependence, whereas $\Delta_{\rm P}$ sets its normalization. Holding the calibrated amplitude $\Delta_{\rm P}$ fixed during the perturbation gives
\begin{equation}
V''_{\rm P}(a_0)=F''_0+\frac{F'_0}{a_0}-\frac{F'^2_0}{F_0}
+\frac{n(a_0F'_0+2F_0)}{a_0^2}.
\label{eq:phantom_vpp}
\end{equation}

\begin{figure*}[htp!]
    \centering
    \includegraphics[width=0.98\linewidth]{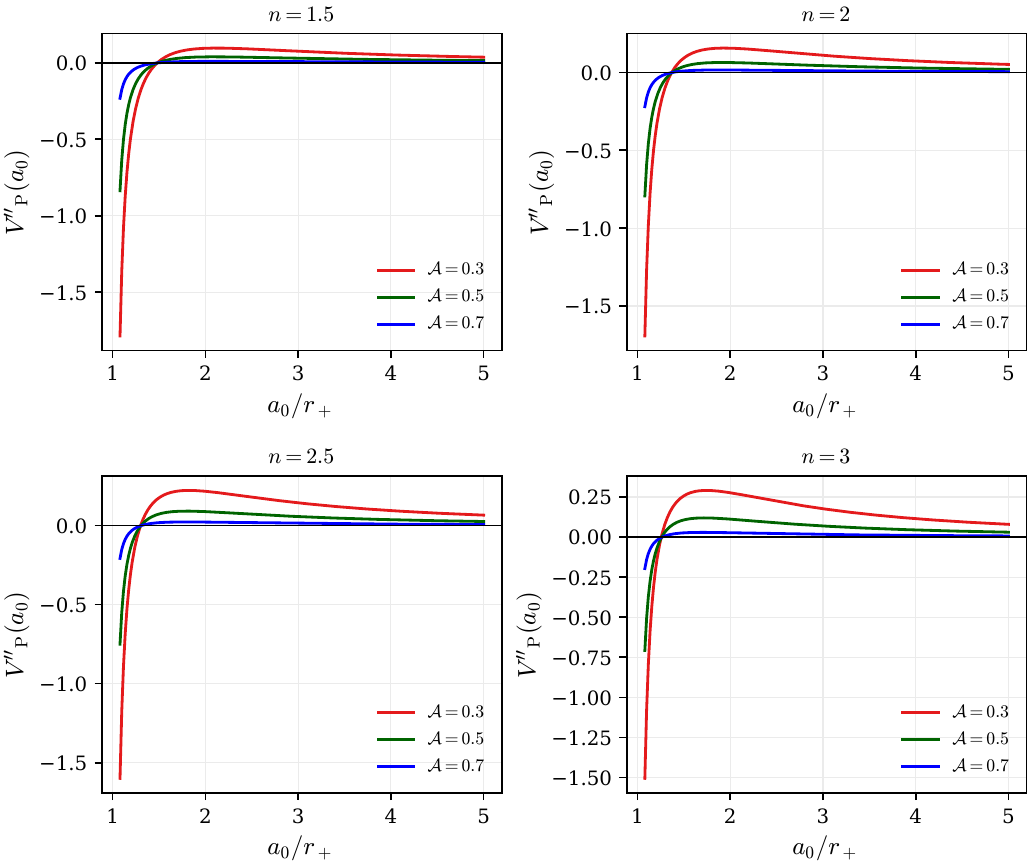}
    \caption{Variable phantom stability function $V''_{\rm P}(a_0)$ as a function of $a_0/r_+$. The panels correspond to $n=1.5,2,2.5,3$, while the curves in each panel correspond to $\mathcal A=0.3,0.5,0.7$. In all panels, $M=1$ and $r_0/M=0.20$; the amplitude $\Delta_{\rm P}$ is fixed by the static junction condition at each displayed throat and held constant during the local perturbation. Only the admissible positive-lapse exterior is plotted. The horizontal zero level separates locally stable configurations, $V''>0$, from unstable ones, $V''<0$.}
    \label{fig:stability_variable_phantom}
\end{figure*}

The explicit radial dependence of the variable phantom law changes the sign structure found for the constant barotropic fluid. In every panel of Fig.~\ref{fig:stability_variable_phantom}, the configurations remain unstable in a narrow region adjacent to the outer horizon, but each curve crosses the zero level and develops a positive branch at larger radii. Increasing $n$ moves this transition toward the horizon and raises the positive maximum, thereby enlarging the locally stable portion of the displayed family. The curves subsequently decrease toward zero from above, with no second sign change visible up to $a_0/r_+=5$.

This behavior is consistent with Eq.~\eqref{eq:phantom_vpp}. The additional term proportional to $n$ is finite at a simple horizon and therefore cannot cancel the negative divergence inherited from the barotropic part, which explains the persistent inner unstable segment. At large radius, however, its leading contribution is proportional to $2n(1-\mathcal A)/a_0^2$ and dominates the faster-decaying negative barotropic terms, producing the positive outer tail. Varying $\mathcal A$ does not act as a purely stabilizing or destabilizing change: larger string strength weakens the negative curvature near the horizon but also suppresses the positive curvature farther out, moving the system toward marginality from both sides. As in the barotropic case, each displayed radius labels a separately calibrated static shell through Eq.~\eqref{eq:phantom_calibration}.

The unregularized limit retains the same mechanism. Equation~\eqref{eq:phantom_vpp} becomes
\begin{equation}
V''_{\mathrm{P,u}}(a_0)
=\frac{(1-\mathcal A)^3}{4M^2x_{\mathrm{u}}^3}
\left[n(2x_{\mathrm{u}}-1)-\frac{x_{\mathrm{u}}}{x_{\mathrm{u}}-1}\right].
\label{eq:phantom_vpp_unregularized}
\end{equation}
The prefactor is positive in the adopted domain. For the positive exponents considered here, the bracket is negative immediately outside the horizon, changes sign once, and becomes positive farther out. Thus the inner unstable and outer stable branches do not rely on central regularization; they originate from the competition between the universal near-horizon term and the explicit radial response of the equation of state. The regulator shifts this balance but does not alter its qualitative topology along the sampled families.

The Bardeen--AdS and Hayward--Letelier--AdS studies use the same radius-dependent linear closure and likewise determine its amplitude from the equilibrium condition \cite{WaseemEtAl2025StringFluid,HouedokouEtAl2026HaywardLetelier}. Their reported dependence on $n$ is not uniform across the two seeds: smaller exponents enlarge the stable domains in the Bardeen slices, whereas higher exponents support stability in the Hayward slices. The present trend agrees with the latter within the displayed family, but Eq.~\eqref{eq:phantom_vpp_unregularized} supplies a stronger, background-specific statement: for the exact Letelier exterior, increasing positive $n$ moves the unique stability transition inward. Thus the exponent acts through its competition with the local lapse terms and is not a geometry-independent stabilizing parameter.

For $n=0$, Eqs.~\eqref{eq:phantom_calibration} and \eqref{eq:phantom_vpp} reduce exactly to Eqs.~\eqref{eq:barotropic_calibration} and \eqref{eq:barotropic_vpp}, respectively.

\subsection{Variable Chaplygin gas}\label{subsec:variable_chaplygin}

The variable Chaplygin closure considered in Refs.~\cite{Varela2015,Javed2024} is
\begin{equation}
p=\frac{\Omega_{\rm C}}{a^n\sigma}.
\label{eq:variable_chaplygin_eos}
\end{equation}
Here $n$ is dimensionless and $[\Omega_{\rm C}]=L^{n-2}$. When $n=0$, the usual inverse-density Chaplygin form is recovered. In the convention often used for a negative thin-shell density, $p=A_{\rm Ch}/|\sigma|$ with $A_{\rm Ch}>0$, this limit corresponds to $\Omega_{\rm C}=-A_{\rm Ch}$. The model is therefore a shell analogue of the Chaplygin gas, not a direct identification with a positive volumetric density.

For Eq.~\eqref{eq:variable_chaplygin_eos},
\begin{equation}
\zeta_{\rm C}=-\frac{p}{\sigma},
\qquad
p_a^{\rm(C)}=-\frac{n}{a}p.
\end{equation}
The static junction condition fixes
\begin{equation}
\Omega_{\rm C}=a_0^n\sigma_0p_0
=-\frac{a_0^{n-2}(a_0F'_0+2F_0)}{16\pi^2}.
\end{equation}
The resulting local stability indicator is
\begin{equation}
V''_{\rm C}(a_0)=F''_0
+\frac{(n+1)F'_0}{a_0}
+\frac{(2n-4)F_0}{a_0^2}.
\label{eq:chaplygin_vpp}
\end{equation}

\begin{figure*}[htp!]
    \centering
    \includegraphics[width=0.98\linewidth]{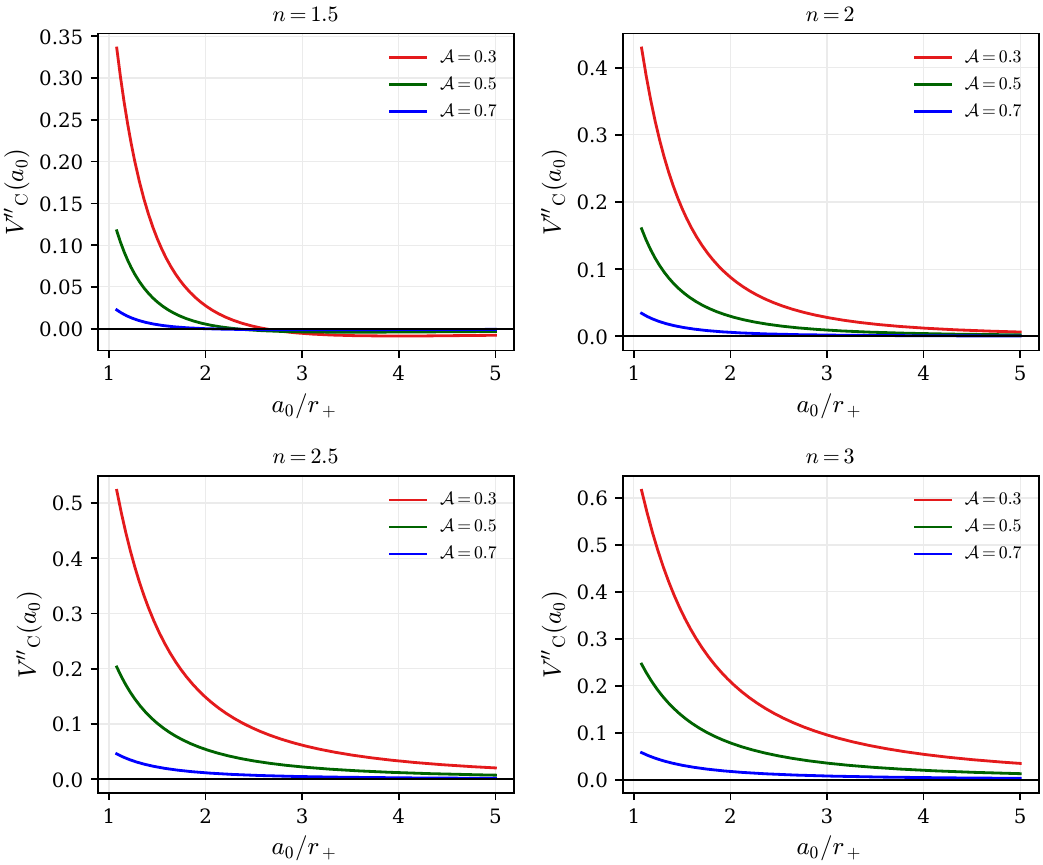}
    \caption{Variable Chaplygin-gas stability function $V''_{\rm C}(a_0)$ as a function of $a_0/r_+$. The panels correspond to $n=1.5,2,2.5,3$, while the curves in each panel correspond to $\mathcal A=0.3,0.5,0.7$. In all panels, $M=1$ and $r_0/M=0.20$; the amplitude $\Omega_{\rm C}$ is fixed by the static junction condition at each displayed throat and held constant during the local perturbation. Only the admissible positive-lapse exterior is plotted. The horizontal zero level separates locally stable configurations, $V''>0$, from unstable ones, $V''<0$.}
    \label{fig:stability_variable_chaplygin}
\end{figure*}

Figure~\ref{fig:stability_variable_chaplygin} displays a qualitatively different near-horizon response. All sampled branches begin with $V''_{\rm C}>0$, so the static configurations immediately outside the plotted lower boundary satisfy the local stability criterion. For $n=1.5$, this positive branch terminates at a single zero and becomes weakly negative farther out; the transition occurs earlier as $\mathcal A$ is increased. By contrast, the curves for $n=2$, $2.5$, and $3$ remain positive throughout the displayed interval, although they approach zero and hence become progressively closer to marginal stability at large radius.

The distinction is controlled by the structure of Eq.~\eqref{eq:chaplygin_vpp}. Unlike Eqs.~\eqref{eq:barotropic_vpp} and \eqref{eq:phantom_vpp}, it contains no term proportional to $-F_0'^2/F_0$, so its simple-horizon limit can remain finite and positive. At large radius, the coefficient $(2n-4)$ separates the behaviors: for $n<2$ the lapse term produces a negative asymptotic tendency, while for $n>2$ it supplies a positive contribution; $n=2$ is the corresponding threshold, with the derivative terms remaining positive for the parameter sets shown. Increasing $\mathcal A$ lowers the potential curvature in every panel and narrows the stable interval when $n=1.5$. The exact horizon is nevertheless excluded, and local stability near it does not remove the pressure-dominated character of the limiting shell.

Here the agreement with the unregularized exterior is also analytic. Equation~\eqref{eq:chaplygin_vpp} becomes
\begin{equation}
V''_{\mathrm{C,u}}(a_0)
=\frac{2\left[(n-2)(1-\mathcal A)a_0+(3-n)M\right]}{a_0^3}.
\label{eq:chaplygin_vpp_unregularized}
\end{equation}
This expression changes sign once for the sampled exponent below two, but remains positive throughout the exterior for the sampled exponents at or above two. Over the interval shown, the regularized profiles exhibit the same separation between a finite inner stable interval and branches that remain stable for the sampled exponents at or above two, with the regulator modifying the transition quantitatively rather than generating the pattern.

A comparable distinction appears in the regular string-fluid studies because both adopt the same variable Chaplygin form and fix its amplitude at equilibrium \cite{WaseemEtAl2025StringFluid,HouedokouEtAl2026HaywardLetelier}. The Hayward--Letelier--AdS analysis reports a transition toward stable configurations at higher $n$, consistent with the separation seen here, whereas the Bardeen--AdS parameter slices favor smaller $n$. Equation~\eqref{eq:chaplygin_vpp_unregularized} resolves the present case without transferring either numerical trend: the threshold at $n=2$ follows analytically from the Letelier lapse and the adopted calibration. The direction of the exponent effect is therefore a joint property of the surface closure and the seed geometry, not of the Chaplygin label alone.

\subsection{Generalized cosmic Chaplygin gas}\label{subsec:gccg}

The generalized cosmic Chaplygin gas (GCCG) was introduced by Gonz\'alez-D\'iaz as a cosmological fluid designed to interpolate between Chaplygin-type behavior and a phantom regime without automatically producing a finite-time big rip \cite{GonzalezDiaz2003}. Its original equation uses a positive volumetric density. Sharif and Mumtaz later applied the same algebraic form to thin-shell matter \cite{SharifMumtaz2014}. Because a surface density has different dimensions and is negative in the present symmetric construction, a fixed positive reference scale $\sigma_\star$ with dimensions $L^{-1}$ is indispensable. We define $x=\sigma/\sigma_\star$ and write
\begin{equation}
\frac{p}{\sigma_\star}
=-x^{-\gamma}
\left[E_{\rm G}+\left(x^{1+\gamma}-E_{\rm G}\right)^{-\omega}\right],
\qquad
E_{\rm G}=\frac{B_{\rm G}}{1+\omega}-1.
\label{eq:gccg_eos}
\end{equation}
Here $\gamma>0$ is a dimensionless Chaplygin exponent, $\omega$ is a dimensionless nonlinear exponent, and $B_{\rm G}\in\mathbb R$ and $E_{\rm G}\in\mathbb R$ are dimensionless parameters related by Eq.~\eqref{eq:gccg_eos}. The reference density satisfies $\sigma_\star>0$ and has dimensions $L^{-1}$, so $x$ is dimensionless. The original GCCG domain is $-\ell<\omega<0$, where $\ell$ is a positive dimensionless bound that may exceed unity; the direct parametrization additionally requires $\omega\neq-1$. We adopt $\ell>10$ so that the sampled values $\omega=-2$ and $-10$ lie in this interval, and we sample $\gamma=1/5$, $3/5$, and $1$, with $\sigma_\star=M^{-1}$. The parameter $B_{\rm G}$ is not scanned independently: its real value, and hence its sign, is recalibrated at every displayed equilibrium. At a static throat, we denote $x(a_0)=\sigma_0/\sigma_\star$ by $x_0$ and the equilibrium-calibrated value of $E_{\rm G}$ by $E_{{\rm G}0}$; the latter is then held fixed under the local perturbation of that particular equilibrium.

The surface application requires an explicit real-power prescription because $x<0$. We restrict $\gamma$ to rational values with an odd denominator in lowest terms, which includes all three sampled values, and evaluate the real powers of $x$ on that branch. Independently of whether the sampled value of $-\omega$ is an integer, the numerical implementation imposes the uniform admissibility condition $x^{1+\gamma}-E_{\rm G}>0$ throughout the local domain; at equilibrium this is $x_0^{1+\gamma}-E_{{\rm G}0}>0$. For the integer values $-\omega=2$ and $10$, a negative base could make the isolated algebraic power real, but such alternatives are deliberately outside the adopted positive-base branch and are not included in the figures. Arbitrary irrational $\gamma$ is likewise excluded unless a different real-valued extension is specified.

The pressure response is
\begin{equation}
\zeta_{\rm G}
=\omega(1+\gamma)
\left(x^{1+\gamma}-E_{\rm G}\right)^{-1-\omega}
-\gamma\frac{p}{\sigma},
\qquad p_a^{\rm(G)}=0.
\end{equation}
For fixed $(\gamma,\omega,\sigma_\star)$, the static junction condition determines a real branch $E_{{\rm G}0}$ through
\begin{equation}
\frac{p_0}{\sigma_\star}
+x_0^{-\gamma}
\left[E_{{\rm G}0}+\left(x_0^{1+\gamma}-E_{{\rm G}0}\right)^{-\omega}\right]=0,
\label{eq:gccg_calibration}
\end{equation}
followed by $B_{\rm G}=(1+\omega)(E_{{\rm G}0}+1)$. For the numerical continuation of the real calibration branch, it is convenient to introduce
\begin{equation}
y=x_0^{1+\gamma}-E_{{\rm G}0}.
\label{eq:gccg_auxiliary_y}
\end{equation}
Here, $y$ is an auxiliary variable representing the argument of the power $-\omega$ in Eq.~\eqref{eq:gccg_calibration}. The adopted branch requires $y>0$ for every sampled value of $\omega$. Using $E_{{\rm G}0}=x_0^{1+\gamma}-y$, the calibration equation becomes
\begin{equation}
y-y^{-\omega}
=
x_0^\gamma
\left(
x_0+\frac{p_0}{\sigma_\star}
\right).
\label{eq:gccg_y_calibration}
\end{equation}
Equation~\eqref{eq:gccg_y_calibration} is generally implicit in $y$. Although $y=1$ would be an exact root at any equilibrium satisfying $p_0=-\sigma_0$, the plotted branch is not initialized at such a reference throat. For each curve, the algorithm starts at the smallest sampled value of $a_0/r_+$ and returns the positive real solution closest to unity, equivalently the retained candidate minimizing $|\ln y|$. At every subsequent radius it selects, from the candidates returned by the solver, the admissible root minimizing $|\ln(y/y_{\rm prev})|$, where $y_{\rm prev}$ is the value retained at the preceding sampled radius. A point is omitted if no returned root satisfies the stated real-power and positive-base conditions. This prescription reproducibly labels a continuous numerical branch; it neither exhausts every possible real-valued extension nor establishes that the retained branch is unique or physically preferred. Once $y$ is determined, the equation-of-state parameters follow from $E_{{\rm G}0}=x_0^{1+\gamma}-y$ and $B_{\rm G}=(1+\omega)(E_{{\rm G}0}+1)$. The calibrated $E_{{\rm G}0}$ is held fixed under local perturbations of each equilibrium configuration.

The second derivative of the potential is
\begin{equation}
\begin{aligned}
V''_{\rm G}(a_0)=&F''_0-\frac{F'^2_0}{2F_0}
+\Bigg[1-2\gamma\frac{p_0}{\sigma_0}
+2\omega(1+\gamma)
\left(x_0^{1+\gamma}-E_{{\rm G}0}\right)^{-1-\omega}\Bigg]
\left(\frac{F'_0}{a_0}-\frac{2F_0}{a_0^2}\right).
\end{aligned}
\label{eq:gccg_vpp}
\end{equation}

\begin{figure*}[htp!]
    \centering
    \includegraphics[width=0.98\linewidth]{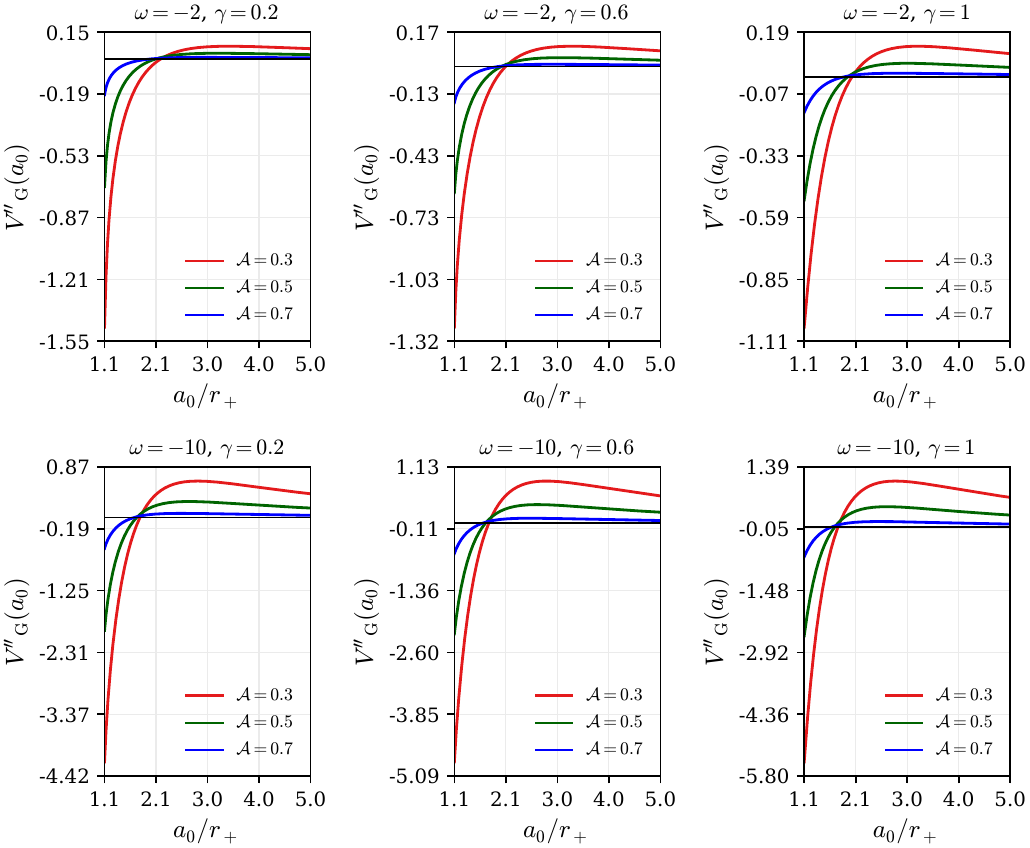}
    \caption{Generalized cosmic Chaplygin-gas stability function $V''_{\rm G}(a_0)$ as a function of the dimensionless throat radius $a_0/r_+$. The columns correspond to $\gamma=1/5$, $3/5$, and $1$, from left to right, while the first and second rows correspond to $\omega=-2$ and $\omega=-10$, respectively. In each panel, the curves correspond to $\mathcal{A}=0.3$, $0.5$, and $0.7$. In all panels, $M=1$, $r_0/M=0.20$, and $\sigma_\star=M^{-1}$; the parameter $B_{\rm G}$ is recalibrated by the static junction condition at each displayed throat. Only points in the admissible positive-lapse exterior that are retained by the positive-real branch-continuation prescription stated in the text are shown. The horizontal zero level separates locally stable configurations, $V''_{\rm G}(a_0)>0$, from unstable ones, $V''_{\rm G}(a_0)<0$.}
    \label{fig:stability_gccg}
\end{figure*}

Across all six panels of Fig.~\ref{fig:stability_gccg}, the selected real branch exhibits the same qualitative sign sequence. The potential curvature is negative at the inner edge of the sampled exterior, crosses the zero level once, and becomes positive at a finite separation from the outer horizon. The positive branch reaches a broad maximum and remains above zero at the largest radius displayed. The inner behavior is consistent with the term $-F_0'^2/(2F_0)$ in Eq.~\eqref{eq:gccg_vpp}, which becomes strongly negative as a simple horizon is approached. Farther outward, the calibrated Chaplygin response can overcome this geometrical contribution, producing a finite interval of locally stable equilibria. This classification concerns spherically symmetric linear perturbations only; it neither establishes global stability nor turns the sequence of displayed radii into the time evolution of a single shell.

Changing $\omega$ from $-2$ to $-10$ preserves this sign topology but substantially increases the magnitude of both the negative and positive branches, while the positions of the zero crossings change comparatively little on the scale shown. The dependence on $\gamma$ is not uniform between the two rows. For $\omega=-2$, increasing $\gamma$ makes the inner negative branch shallower and raises the outer positive maximum. For $\omega=-10$, the same variation deepens the negative branch and also increases the positive maximum, thereby enhancing the contrast between the two regimes. These trends cannot be assigned to the explicit $\gamma$- and $\omega$-dependent factors alone, because $E_{{\rm G}0}$, and hence $B_{\rm G}$, is recalibrated through Eq.~\eqref{eq:gccg_calibration} at every static radius.

Within each panel, increasing the string strength $\mathcal A$ brings the curves closer to the zero level and shifts the stability transition toward smaller $a_0/r_+$. The larger-$\mathcal A$ configurations therefore enter the positive branch closer to the horizon in the normalized coordinate, but their positive potential curvature is also weaker, so the net effect is better described as a displacement toward marginality rather than as a uniformly stabilizing change. Each plotted point represents a separately calibrated GCCG-supported static throat on the continuous positive real branch. The analytic reductions of the model remain consistent with this construction: the $\omega\rightarrow0$ limit recovers the generalized Chaplygin law, and the additional choice $\gamma=1$ gives the ordinary Chaplygin case.

For the GCCG, removing the regulator makes every geometric coefficient in Eq.~\eqref{eq:gccg_vpp} elementary, but the calibrated quantity $E_{{\rm G}0}$ remains tied to the implicit branch equation~\eqref{eq:gccg_y_calibration}. Consequently, the unregularized comparison is semi-analytic rather than a universal closed-form sign proof. As long as the selected positive real root remains simple and admissible, implicit-function continuity carries the same calibrated branch smoothly to $r_0\rightarrow0$. Continuity of the branch variables, however, does not preserve the sign of $V''_{\rm G}$: the stability function can cross zero without any loss or merger of its calibration root. The displayed inner-unstable/outer-stable topology therefore characterizes the sampled regularized family only. Establishing that it persists at $r_0=0$ requires following both the root and the zeros of $V''_{\rm G}$ throughout the continuation, so this topology should not be attributed either to regularity or to its absence from branch continuity alone.

\subsection{Modified cosmic Chaplygin gas}\label{subsec:mccg}

Sadeghi and Farahani considered a varying, viscous modified cosmic Chaplygin gas (MCCG) in a cosmological setting, with a scale-factor-dependent coefficient and a time-dependent cosmic parameter \cite{SadeghiFarahani2013}. The thin-shell model used by Sharif and Mumtaz is the constant-parameter, nonviscous specialization of that broader construction \cite{SharifMumtaz2014}. With the same reference scale and branch convention used above, its surface form is
\begin{equation}
\frac{p}{\sigma_\star}
=A_{\rm M}x
-x^{-\gamma}
\left[E_{\rm M}+\left(x^{1+\gamma}-E_{\rm M}\right)^{-\omega}\right],
\qquad
E_{\rm M}=\frac{B_{\rm M}}{1+\omega}-1.
\end{equation}
In the adopted surface closure, $A_{\rm M}\in[0,\infty)$ is a dimensionless coefficient of the linear term, while $B_{\rm M}\in\mathbb R$ and $E_{\rm M}\in\mathbb R$ are distinct MCCG parameters related by the equation above. This non-negative domain for $A_{\rm M}$ is an explicit restriction of our numerical implementation, not a cosmological inference. We fix $A_{\rm M}=2$ solely as a representative value used in the thin-shell comparison of Ref.~\cite{SharifMumtaz2014}; it is neither fitted nor claimed to be theoretically preferred. The remaining adopted domain is $\gamma>0$, $-\ell<\omega<0$ with $\omega\neq-1$ and $\ell>10$, together with the real-power and positive-base conditions stated for the GCCG. The figures sample $\gamma=1/5$, $3/5$, and $1$, $\omega=-2$ and $-10$, and $\sigma_\star=M^{-1}$. The parameter $B_{\rm M}$ is not scanned independently, because its real value is recalibrated at each static throat. At equilibrium, $E_{{\rm M}0}$ denotes that calibrated value of $E_{\rm M}$ and is held fixed during the corresponding local perturbation.

The equation-of-state derivatives are
\begin{equation}
\begin{aligned}
\zeta_{\rm M}
&=A_{\rm M}(1+\gamma)-\gamma\frac{p}{\sigma}
+\omega(1+\gamma)
\left(x^{1+\gamma}-E_{\rm M}\right)^{-1-\omega},
& p_a^{\rm(M)}&=0.
\end{aligned}
\end{equation}
For fixed $(A_{\rm M},\gamma,\omega,\sigma_\star)$, the static condition fixes $E_{{\rm M}0}$ implicitly:
\begin{equation}
\frac{p_0}{\sigma_\star}-A_{\rm M}x_0
+x_0^{-\gamma}
\left[E_{{\rm M}0}+\left(x_0^{1+\gamma}-E_{{\rm M}0}\right)^{-\omega}\right]=0.
\label{eq:mccg_calibration}
\end{equation}
Using the same auxiliary branch variable introduced in the GCCG case, \eqref{eq:gccg_auxiliary_y}, the MCCG calibration can be written as
\begin{equation}
y-y^{-\omega}
=
x_0^\gamma
\left[
\frac{p_0}{\sigma_\star}
+
(1-A_{\rm M})x_0
\right].
\label{eq:mccg_y_calibration}
\end{equation}
Thus $y$ again denotes the argument of the nonlinear Chaplygin power, now with $y=x_0^{1+\gamma}-E_{{\rm M}0}$. The MCCG curves use the same numerical selection rule as the GCCG curves: at the smallest sampled radius the positive root nearest $y=1$ is retained, and subsequent admissible roots are selected by logarithmic proximity to the preceding value. Although $y=1$ would be exact when $p_0/\sigma_\star+(1-A_{\rm M})x_0=0$, no sampled reference throat satisfying that relation is assumed. The rule labels the branch followed by the calculation but does not prove its uniqueness or physical preference. Once this root is chosen, $E_{{\rm M}0}=x_0^{1+\gamma}-y$ and $B_{\rm M}=(1+\omega)(E_{{\rm M}0}+1)$. As in the GCCG case, only real roots satisfying the local branch conditions are admissible, and the selected $E_{{\rm M}0}$ is held fixed during the perturbation.

The corresponding stability expression is
\begin{equation}
\begin{aligned}
V''_{\rm M}(a_0)=&F''_0-\frac{F'^2_0}{2F_0}\\
&+\Bigg[1+2A_{\rm M}(1+\gamma)-2\gamma\frac{p_0}{\sigma_0}
+2\omega(1+\gamma)
\left(x_0^{1+\gamma}-E_{{\rm M}0}\right)^{-1-\omega}\Bigg]
\left(\frac{F'_0}{a_0}-\frac{2F_0}{a_0^2}\right).
\end{aligned}
\end{equation}
Setting $A_{\rm M}=0$ and identifying $E_{\rm M}=E_{\rm G}$, equivalently $B_{\rm M}=B_{\rm G}$, recovers Eqs.~\eqref{eq:gccg_eos}, \eqref{eq:gccg_calibration}, and \eqref{eq:gccg_vpp}. Taking $\omega\rightarrow0$ gives the modified generalized Chaplygin law $p/\sigma_\star=A_{\rm M}x-B_{\rm M}x^{-\gamma}$. The further choice $\gamma=1$ yields the modified Chaplygin gas, while $A_{\rm M}=0$ then gives the ordinary Chaplygin limit. These reductions are algebraic and do not require importing the cosmological evolution, viscosity, or scale-factor dependence of the original varying model.

\begin{figure*}[htp!]
    \centering
    \includegraphics[width=0.98\linewidth]{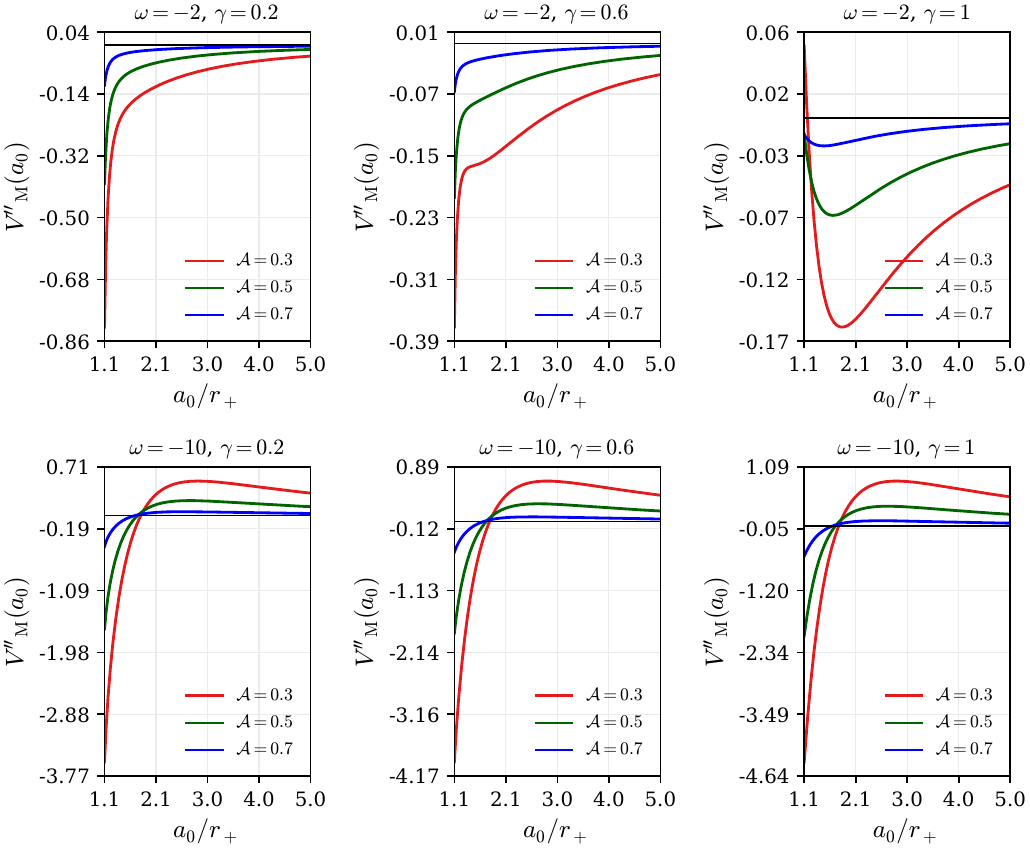}
    \caption{Modified cosmic Chaplygin-gas stability function $V''_{\rm M}(a_0)$ as a function of the dimensionless throat radius $a_0/r_+$. The columns correspond to $\gamma=1/5$, $3/5$, and $1$, from left to right, while the first and second rows correspond to $\omega=-2$ and $\omega=-10$, respectively. In each panel, the curves correspond to $\mathcal{A}=0.3$, $0.5$, and $0.7$. In all panels, $M=1$, $r_0/M=0.20$, $A_{\rm M}=2$, and $\sigma_\star=M^{-1}$; the parameter $B_{\rm M}$ is recalibrated by the static junction condition at each displayed throat. Only points in the admissible positive-lapse exterior that are retained by the positive-real branch-continuation prescription stated in the text are shown. The horizontal zero level separates locally stable configurations, $V''_{\rm M}(a_0)>0$, from unstable ones, $V''_{\rm M}(a_0)<0$.}
    \label{fig:stability_mccg}
\end{figure*}

Figure~\ref{fig:stability_mccg} reveals a marked dependence of the MCCG stability profiles on $\omega$. For $\omega=-2$, no systematic outer stable branch is found. At $\gamma=0.2$ and $0.6$, all three curves remain below the zero level throughout the displayed interval, although increasing the string-cloud parameter $\mathcal A$ brings them progressively closer to marginality. The behavior at $\gamma=1$ is nonmonotonic: the $\mathcal A=0.3$ branch is positive only over a short segment adjacent to the lower plotted boundary, crosses downward, develops the deepest negative minimum, and remains negative thereafter. The $\mathcal A=0.5$ and $0.7$ branches remain below zero within the resolved range. Since the plot begins at $a_0/r_+=1.1$ rather than at the horizon itself, the narrow positive segment of the $\mathcal A=0.3$ curve should not be interpreted as establishing stability in the exact horizon limit.

A qualitatively more uniform pattern is obtained for $\omega=-10$. Every branch in the lower row begins with negative potential curvature, crosses the zero level once, and develops a positive outer branch that persists up to the largest radius shown. At fixed $\gamma$, increasing $\mathcal A$ reduces the magnitudes of both the negative inner branch and the positive outer maximum, while shifting the zero crossing toward smaller $a_0/r_+$. At fixed $\mathcal A$, increasing $\gamma$ deepens the negative minimum and raises the positive maximum. Thus, within the sampled parameter set, the string-cloud strength mainly compresses the profiles toward marginality, whereas $\gamma$ enhances the contrast between the unstable and stable radial regimes when $\omega=-10$.

The comparison with Fig.~\ref{fig:stability_gccg} shows that the MCCG modification is not equivalent to a uniform rescaling of the GCCG curves. For $\omega=-2$, the outer positive branches present in the GCCG case are absent from almost all MCCG profiles, whereas for $\omega=-10$ the GCCG sign topology is recovered but with generally smaller amplitudes. According to the MCCG stability expression, the explicit linear contribution is proportional to $A_{\rm M}(1+\gamma)(F'_0/a_0-2F_0/a_0^2)$. This geometrical factor changes sign at the same radius that controls the null-energy-condition transition discussed in Fig.~\ref{fig:static_energy_conditions}; consequently, the linear term raises the potential curvature on one side of that transition and lowers it on the other. Nevertheless, the difference between the GCCG and MCCG profiles cannot be attributed to this explicit term alone, because the static condition also changes the calibrated value of $E_{{\rm M}0}$, and hence of $B_{\rm M}$.

In each panel, $A_{\rm M}$ is fixed at $2$, so the separation among the three curves results from varying the string-cloud parameter $\mathcal A$, through both the seed lapse and the recalibrated MCCG parameter. The change from $\omega=-2$ to $\omega=-10$ produces the clearest qualitative reorganization within this numerical scan, but this conclusion is restricted to the selected continuous positive real branch and to the parameter values displayed. Moreover, because $B_{\rm M}$ is recalibrated at each throat, the curves represent families of distinct static equilibria rather than the radial evolution of a single shell.

The same qualification applies to the MCCG comparison. At $r_0=0$, the lapse-dependent terms become elementary, whereas $E_{{\rm M}0}$ must still be followed on the selected solution of Eq.~\eqref{eq:mccg_y_calibration}. Continuity of the calibrated quantities again guarantees continuity of $V''_{\rm M}$, not conservation of its sign. An explicit continuation of the sampled $\mathcal A=0.3$, $\gamma=1$, $\omega=-2$, $A_{\rm M}=2$ configuration at $a_0/r_+=1.1$ illustrates the distinction: its potential curvature is positive at $r_0/M=0.20$ but negative at $r_0=0$, while the selected positive calibration root remains real and simple. Hence the continuation crosses marginality without losing the equation-of-state branch. This fixed-$(M,\mathcal A,a_0/r_+)$ comparison does not by itself isolate a core-regularity effect, because the physical throat radius, horizon scale, and asymptotic mass coefficient also change. Because both the explicit linear term and the recalibrated nonlinear parameter contribute, the strong $\omega$ dependence in Fig.~\ref{fig:stability_mccg} belongs to the shell closure and its branch selection and cannot be promoted to a universal effect of smoothing the central geometry.

In addition, all five equations of state are used here as phenomenological closures for distributional shell matter. Cosmological interpretations attached to volumetric fluids cannot be transferred automatically to the surface variables $\sigma$ and $p$, which both have dimensions $L^{-1}$ in geometrized units.

\section{Optical appearance of the thin-shell wormhole}\label{sec:optical_appearance}
\subsection{Thin-disk model}\label{subsec:thin_disk}

We place a geometrically thin emitting disk on the equatorial plane of the observer's exterior. The accretion flow is modeled by equatorial circular timelike geodesics with four-velocity
\begin{equation}
u^\mu=u^t(1,0,0,\Omega),
\qquad
\Omega^2=\frac{F'(r)}{2r},
\qquad
u^t=\left[F(r)-r^2\Omega^2\right]^{-1/2}
=\sqrt{\frac{2}{2F(r)-rF'(r)}}.
\label{eq:disk_four_velocity}
\end{equation}
A circular timelike orbit requires $F'(r)>0$ and $2F(r)-rF'(r)>0$. Its marginal-stability radius is determined by
\begin{equation}
rF(r)F''(r)+3F(r)F'(r)-2rF'(r)^2=0.
\label{eq:isco_condition}
\end{equation}
We denote by $r_{\rm ISCO}$ the innermost root that bounds the outer stable branch. For the imaging configurations considered below, we restrict the throat to $a_0<r_{\rm ISCO}$ and place the inner edge of the emitting disk at
\begin{equation}
r_{\rm in}=r_{\rm ISCO}.
\label{eq:disk_inner_edge}
\end{equation}
This restriction is necessary because an ISCO lying below the throat would belong to the excised region and could not define a physical disk edge in the glued spacetime.

For the radial emissivity we adopt the phenomenological GLM profile \cite{grallaShapeBlackHole2020} with $\gamma_{\rm em}=0$, $\mu_{\rm em}=r_{\rm ISCO}$, and $\sigma_{\rm em}=M$,
\begin{equation}
I_{\nu}(r)=
\frac{\exp\!\left[-\frac{1}{2}
\operatorname{arcsinh}^2\!\left(\frac{r-r_{\rm ISCO}}{M}\right)\right]}
{\sqrt{(r-r_{\rm ISCO})^2+M^2}}
\,\Theta(r-r_{\rm ISCO}).
\label{eq:emission_profile}
\end{equation}
The step function removes all emission for $r<r_{\rm ISCO}$, while the remaining profile reaches its maximum at the inner edge. No outer cutoff is imposed; the emissivity decreases smoothly as $r\rightarrow\infty$. This choice defines an edge-brightened phenomenological disk rather than the zero-torque thin-disk profile, whose emitted flux vanishes at the ISCO.

\subsection{Imaging method}\label{subsec:imaging_method}

We construct images of the thin-shell wormhole surrounded by a geometrically thin accretion disk using backward ray tracing, following the numerical framework described in Ref.~\cite{huQEDEffectBlack2021}. Photon trajectories are obtained from the Hamiltonian
\begin{equation}
\mathcal H(x,p)=\frac{1}{2}g^{\mu\nu}p_\mu p_\nu=0,
\label{eq:photon_hamiltonian}
\end{equation}
where the last equality imposes the null condition. Hamilton's equations take the first-order form
\begin{equation}
\frac{\dd x^\mu}{\dd\lambda}
=\frac{\partial\mathcal H}{\partial p_\mu}
=g^{\mu\nu}p_\nu,
\qquad
\frac{\dd p_\mu}{\dd\lambda}
=-\frac{\partial\mathcal H}{\partial x^\mu}
=-\frac{1}{2}\,\partial_\mu g^{\alpha\beta}p_\alpha p_\beta,
\label{eq:photon_hamilton_equations}
\end{equation}
where $\lambda$ is an affine parameter.

The observer is placed at $(t_o,r_o,\theta_o,\phi_o)$ in the positive-lapse exterior. Because the spacetime in Eq.~\eqref{eq:seed_metric} is static and spherically symmetric, a zero-angular-momentum observer (ZAMO) coincides with a static observer. A convenient orthonormal tetrad at the observation event is
\begin{equation}
\begin{aligned}
\hat e_{(0)}&=\frac{1}{\sqrt{F_o}}\,\partial_t,
&\qquad
\hat e_{(1)}&=-\sqrt{F_o}\,\partial_r,\\
\hat e_{(2)}&=\frac{1}{r_o}\,\partial_\theta,
&
\hat e_{(3)}&=-\frac{1}{r_o\sin\theta_o}\,\partial_\phi,
\end{aligned}
\label{eq:observer_tetrad}
\end{equation}
where $F_o\equiv F(r_o)$. The signs of the spatial basis vectors fix the orientation of the observer's screen, with $\hat e_{(1)}$ directed toward decreasing areal radius. The tetrad components of the initial photon momentum determine a direction on the local sky. We map these directions to the image plane by stereographic projection and integrate Eqs.~\eqref{eq:photon_hamilton_equations} backward from the observer.

Following Refs.~\cite{wangNovelShadowsAsymmetric2020,langloisConservationLawsCollisions2002}, let $\{\hat e_{t_\pm},\hat e_{r_\pm}\}$ denote the orthonormal frames associated with the two coordinate charts at the throat. The frames are related by
\begin{equation}
\begin{pmatrix}
\hat e_{t_-} \\
\epsilon_-\hat e_{r_-}
\end{pmatrix}
=
\Lambda(\Theta)
\begin{pmatrix}
\hat e_{t_+} \\
\epsilon_+\hat e_{r_+}
\end{pmatrix},
\qquad
\Lambda(\Theta)=
\begin{pmatrix}
\cosh\Theta & \sinh\Theta \\
\sinh\Theta & \cosh\Theta
\end{pmatrix},
\end{equation}
where
\begin{equation}
\Theta=
\sinh^{-1}\!\left[
\frac{\epsilon_+\dot a}{\sqrt{F_+(a)}}
\right]
-
\sinh^{-1}\!\left[
\frac{\epsilon_-\dot a}{\sqrt{F_-(a)}}
\right],
\end{equation}
and $F_\pm$ denotes the lapse function on $\mathcal M^\pm$; in the present symmetric construction, $F_+=F_-=F$. We orient the glued manifold from $\mathcal M^-$ to $\mathcal M^+$, i.e., from left to right. Along this direction, $r_-$ decreases whereas $r_+$ increases. With the orientation convention of Ref.~\cite{langloisConservationLawsCollisions2002}, this gives $\epsilon_-=+1$ and $\epsilon_+=-1$. For a static shell, $\dot a=0$, so that $\Theta=0$ and $\Lambda(0)=I$. The matching relation then reduces to $\hat e_{t_-}=\hat e_{t_+}$ and $\hat e_{r_-}=-\hat e_{r_+}$. The shell is assumed to be optically transparent, without absorption, scattering, or refraction. Under this assumption, the photon crosses the shell without any nongravitational exchange of momentum; continuity of its four-momentum, together with the reversal of the radial frame, therefore implies $p^{\hat r}_+=-p^{\hat r}_-$, so that the radial momentum component changes sign when expressed in the chart on the other side. This sign change represents the transition between the two oppositely oriented radial frames and does not correspond to a physical reflection at the throat.

For all images, the ZAMO and its local screen are placed at $r_o=200M$. We scale the field of view separately for each seed geometry using its photon-sphere impact scale
\begin{equation}
b_{\rm ph}=\frac{r_{\rm ph}}{\sqrt{F(r_{\rm ph})}},
\qquad
r_{\rm ph}F'(r_{\rm ph})-2F(r_{\rm ph})=0,
\label{eq:photon_impact_scale}
\end{equation}
where $r_{\rm ph}$ is the outer photon-sphere radius of the corresponding seed metric. The stereographic image plane covers $[-2.4b_{\rm ph},2.4b_{\rm ph}]^2$, so that $2.4b_{\rm ph}$ is the screen half-width, and is sampled with $2000\times2000$ pixels. This parameter-dependent normalization keeps the same multiple of the photon-sphere scale within the field of view throughout the scan.

\begin{table*}[t]
    \centering
    \caption{Characteristic radial scales of the regularized string-cloud seed geometries used for imaging. Here $M=1$ and $r_0/M=0.20$. The outer event horizon $r_+$, outer photon sphere $r_{\rm ph}$, and ISCO bounding the outer stable circular-orbit branch are obtained from $F(r_+)=0$, Eq.~\eqref{eq:photon_impact_scale}, and Eq.~\eqref{eq:isco_condition}, respectively. The last two columns use the outer-horizon normalization employed for the throat radii.}
    \label{tab:optical_scales}
    \begin{ruledtabular}
    \begin{tabular}{cccccc}
    $\mathcal A$ & $r_+/M$ & $r_{\rm ph}/M$ & $r_{\rm ISCO}/M$ & $r_{\rm ph}/r_+$ & $r_{\rm ISCO}/r_+$ \\
    \hline
    $0.3$ & $1.269$ & $2.132$ & $4.109$ & $1.679$ & $3.237$ \\
    $0.5$ & $1.769$ & $2.864$ & $5.534$ & $1.620$ & $3.129$ \\
    $0.7$ & $2.923$ & $4.582$ & $8.917$ & $1.568$ & $3.051$
    \end{tabular}
    \end{ruledtabular}
\end{table*}

Table~\ref{tab:optical_scales} makes the motivation for the two throat radii explicit. All three string-cloud strengths obey the same ordering

\begin{equation*}
r_+<1.2r_+<r_{\rm ph}<2.0r_+<r_{\rm ISCO}.
\end{equation*}

Thus, the throat at $a_0/r_+=1.2$ lies between the outer horizon and the outer photon sphere and retains the seed photon sphere, whereas the throat at $a_0/r_+=2.0$ lies outside and excises it. Both throats remain below $r_{\rm ISCO}$, so the disk inner edge $r_{\rm in}=r_{\rm ISCO}$ belongs to the retained region in every imaging configuration. The two radii therefore bracket a physically distinguished optical scale rather than constituting an arbitrary radial sample.

The specific intensity measured by the observer is obtained from the covariant radiative-transfer equation
\begin{equation}
\frac{\dd}{\dd\lambda}\left(\frac{I_\nu}{\nu^3}\right)
=\frac{j_\nu-\alpha_\nu I_\nu}{\nu^2},
\label{eq:radiative_transfer}
\end{equation}
where $I_\nu$, $j_\nu$, and $\alpha_\nu$ are, respectively, the specific intensity, emissivity, and absorption coefficient at frequency $\nu$. In the thin-disk model adopted here, the emissivity is confined to the equatorial plane. We neglect absorption, so that $\alpha_\nu=0$ and $I_\nu/\nu^3$ is conserved along each vacuum segment of a ray between successive intersections with the disk. The total observed intensity is therefore
\begin{equation}
I_o=\sum_n g_n^3 I_n,
\label{eq:observed_intensity}
\end{equation}
where $I_n$ is the emitted specific intensity evaluated at the $n$th intersection of the ray with the equatorial plane. The adopted $g^3$ weighting defines a monochromatic specific-intensity prescription rather than a bolometric image. The corresponding redshift factor is
\begin{equation}
g_n\equiv\frac{\nu_o}{\nu_n}
=\frac{\left(p_\mu u_o^\mu\right)_o}
{\left(p_\mu u^\mu\right)_n},
\label{eq:redshift_factor}
\end{equation}
where $u_o^\mu=\hat e_{(0)}^\mu=F_o^{-1/2}(1,0,0,0)$ is the four-velocity of the ZAMO, $u^\mu$ is the four-velocity of the accretion flow, and $p_\mu$ is the photon four-momentum. We normalize the photon momentum by setting $\nu_o=1$, in which case Eq.~\eqref{eq:redshift_factor} reduces to $g_n=-1/(u^\mu p_\mu)|_n$. The index $n$ indicates that all disk quantities are evaluated at the point where the ray crosses the equatorial plane for the $n$th time.

\subsection{Imaging results}\label{subsec:imaging_results}

We first present images of the celestial sphere. Rays that, when traced backward, reach the celestial sphere in the observer's universe are assigned red, green, yellow, and blue to represent its four regions. Rays that cross the throat and reach the celestial sphere in the other universe are instead assigned pink, indigo, purple, and orange. The two color palettes distinguish images originating from the two universes. The white marker denotes the point source on the celestial sphere defined by the line connecting the observer and the wormhole center.

\begin{figure*}[htp!]
    \centering
    \includegraphics[width=0.85\linewidth]{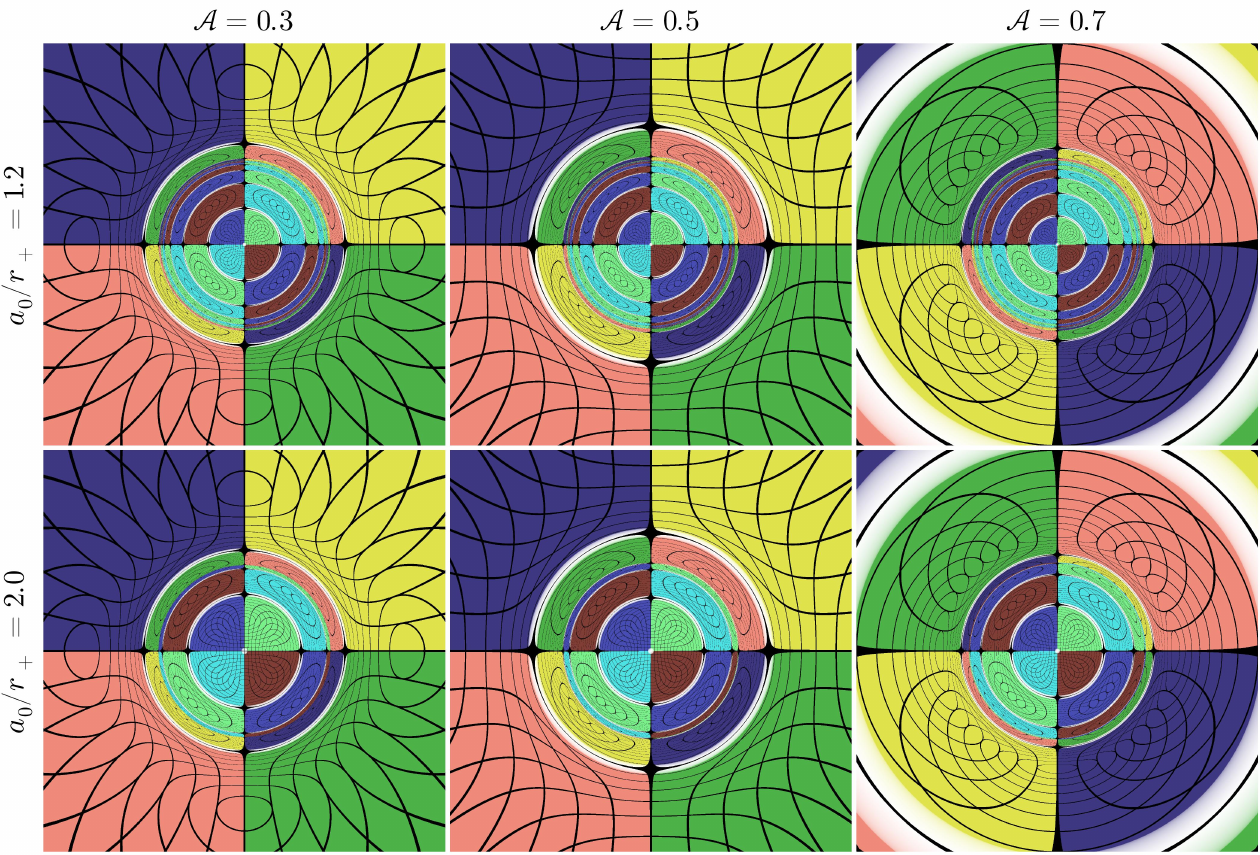}
    \caption{Images of the celestial sphere around the symmetric thin-shell wormhole. From left to right, the columns show $\mathcal A=0.3$, $0.5$, and $0.7$; the upper and lower rows show $a_0/r_+=1.2$ and $2.0$, respectively. The remaining parameters are $M=1$, $r_0/M=0.20$, $r_o=200M$, and $\theta_o=80^\circ$, with screen half-width $2.4b_{\rm ph}$. Colors and black curves mark positions on a diagnostic celestial-sphere pattern and do not represent radiative intensity. Repeated images of the pattern in multiple annuli reveal the multiple Einstein rings and higher-order image structure.}
    \label{fig:wormhole_lensing_images}
\end{figure*}

Because the spacetime is spherically symmetric, the image of the celestial sphere does not depend on the viewing inclination. To separate the effects of the string-cloud strength and the throat location, Fig.~\ref{fig:wormhole_lensing_images} compares three values of $\mathcal A$ at the two dimensionless throat radii motivated by Table~\ref{tab:optical_scales}. The columns correspond to $\mathcal A=0.3$, $0.5$, and $0.7$, whereas the upper and lower rows correspond to $a_0/r_+=1.2$ and $2.0$, respectively. The remaining parameters are held fixed, with the viewing inclination set to $\theta_o=80^\circ$.

In Fig.~\ref{fig:wormhole_lensing_images}, the white point source aligned with the observer and the wormhole center forms multiple Einstein rings under gravitational lensing, with additional multiple-image structures appearing between the rings. Changing the junction radius mainly reorganizes the image branches that probe the excised region. At fixed $\mathcal A$, the outer direct image is nearly unchanged between the two rows because the corresponding rays remain in the exterior geometry common to both configurations. By contrast, moving the gluing surface from $a_0/r_+=1.2$ to $2.0$ excises the region near the photon sphere, thereby reducing the corresponding higher-order image branches and changing which strongly deflected rays can be continued into the second exterior. The smaller throat therefore generally retains a denser set of narrow central replicas, most clearly for $\mathcal A=0.3$ and $0.5$. Increasing $\mathcal A$ also changes the widths and separations of the annuli, making the distinct higher-order image domains easier to identify.

\begin{figure*}[htp!]
    \centering
    \includegraphics[width=0.85\linewidth]{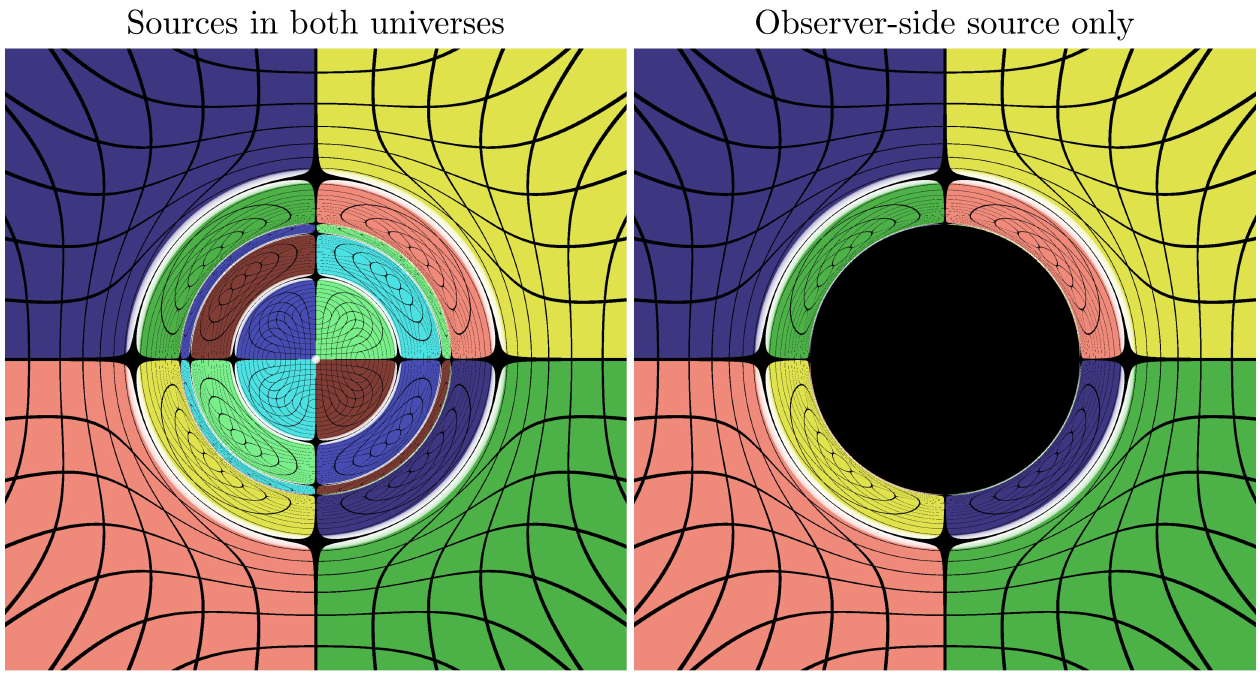}
    \caption{Comparison of identical celestial spheres placed in both universes (left) with a celestial sphere placed only in the observer's universe (right). The two panels use exactly the same null-geodesic data and parameters $(\mathcal A,a_0/r_+,\theta_o)=(0.5,2.0,80^\circ)$, with $M=1$, $r_0/M=0.20$, $r_o=200M$, and screen half-width $2.4b_{\rm ph}$. Colors mark the endpoints of the rays on the celestial-sphere test pattern and do not represent radiative intensity; the black regions in the right panel indicate rays that terminate in the other universe, where no source is placed.}
    \label{fig:wormhole_lensing_onside_source_images}
\end{figure*}

Fig.~\ref{fig:wormhole_lensing_onside_source_images} compares the configuration with identical celestial spheres placed in both universes (left) with the configuration in which the celestial sphere is present only in the observer's universe (right). This comparison changes only the boundary condition of the sources and leaves the wormhole geometry, together with every null geodesic, unchanged. The direct images, the returning images, and the white Einstein ring on the observer's side are therefore identical in the two panels, whereas the inner colored replica images formed by through-throat rays in the left panel become black in the right panel, because those rays now terminate in the empty second exterior. The boundary of this black region is governed by the location of the throat relative to the seed photon sphere. For $a_0<r_{\rm ph}$, the outer photon sphere of the seed spacetime is retained in the glued manifold, and the boundary is controlled by the corresponding critical curve; on the basis of this boundary alone, the wormhole generally cannot be distinguished from the corresponding black hole. For $a_0>r_{\rm ph}$, by contrast, the junction excises the outer photon sphere, and the boundary is instead set by the critical rays that graze the throat. A measured boundary radius exceeding the image-plane critical radius $b_{\rm ph}$ of the outer photon sphere would therefore provide a potential optical criterion for distinguishing the wormhole from its black-hole counterpart. The configuration shown in the figure belongs to the case $a_0>r_{\rm ph}$.

We next include the thin-disk emissivity and redshift. To enhance faint higher-order structures without rescaling each panel independently, we display the raw observed intensity $I_o$ using the nonlinear map
\begin{equation}
I_{\log}=\begin{cases}
-\displaystyle\left[\log_2\!\left(
\dfrac{I_o}{I_{\max}+I_o}
\right)\right]^{-1}, & I_o>0,\\[6pt]
0, & I_o=0,
\end{cases}
\label{eq:log_intensity_normalization}
\end{equation}
where $I_{\max}$ is the largest raw intensity over all pixels and all 18 configurations in the primary scan. The same $I_{\max}$ is used for every panel, so that $0\leq I_{\log}\leq1$ and $I_o=I_{\max}$ maps to $I_{\log}=1$. This common nonlinear normalization increases the visibility of faint lensed structures while preserving intensity differences among the panels; no panel-by-panel renormalization is applied.

\begin{figure*}[htp!]
    \centering
    \includegraphics[width=0.85\linewidth]{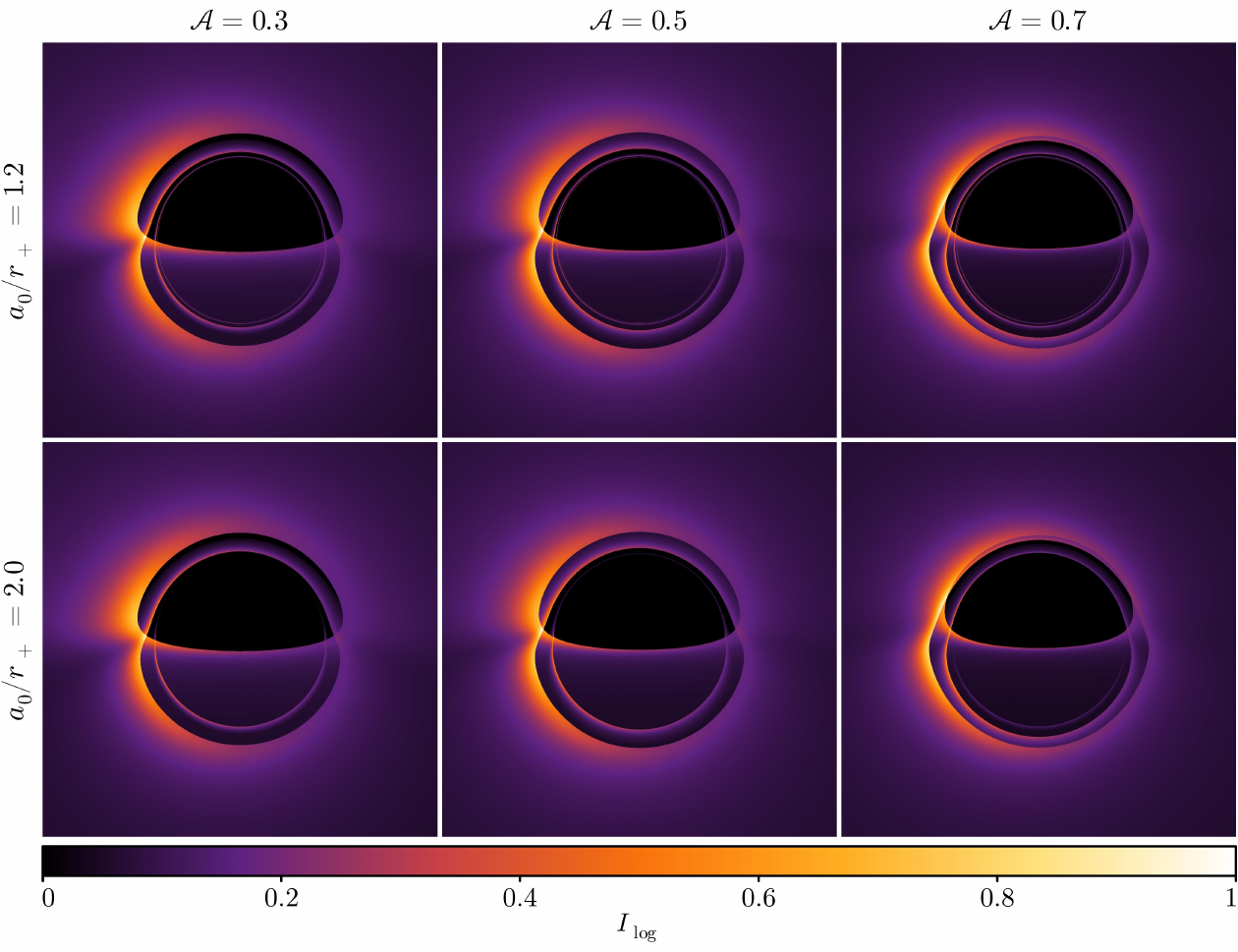}
    \caption{Logarithmically mapped thin-disk images of the symmetric thin-shell wormhole for different string-cloud strengths and throat radii. From left to right, the columns show $\mathcal A=0.3$, $0.5$, and $0.7$; the upper and lower rows show $a_0/r_+=1.2$ and $2.0$, respectively. The remaining parameters are fixed at $M=1$, $r_0/M=0.20$, $r_o=200M$, and $\theta_o=80^\circ$. All panels use the same GLM emissivity profile, the screen half-width $2.4b_{\rm ph}$, and the common normalization defined in Eq.~\eqref{eq:log_intensity_normalization}.}
    \label{fig:wormhole_thin_disk_images}
\end{figure*}

Figure~\ref{fig:wormhole_thin_disk_images} uses the same parameter choices as Fig.~\ref{fig:wormhole_lensing_images} to show how the string-cloud strength and throat radius affect the thin-disk image.

At either throat radius, the nearly edge-on disk consists of a bright direct-image component and a sequence of nested arcs and narrow rings. The strong left--right intensity contrast arises from the orbital redshift factor: the side moving toward the observer is Doppler enhanced, whereas the receding side is suppressed. The thin line crossing the central region is the projected direct image of the disk, while the curved branches above and below it are lensed images of the front and back portions of the disk. In this one-sided emission model, the central dark region is governed mainly by truncating the accretion disk at the ISCO. It therefore should not be identified directly with a black-hole shadow, nor does its presence alone establish that the corresponding rays are captured.

As $\mathcal A$ increases, the initially broad structures split into a more clearly resolved set of higher-order image branches. This effect is especially evident for $\mathcal A=0.7$, where several distinct rings can be identified around the central dark region. These structures are not additional emitting components but multiple images of the same disk. Their dependence on the throat radius follows the scale ordering in Table~\ref{tab:optical_scales}: the smaller throat retains the outer photon sphere and its nearby multi-orbit trajectories, whereas the larger throat excises that region and reduces mainly the inner higher-order branches. Exterior direct rays remain insensitive to the gluing radius because they propagate in the geometry common to both configurations. Because every panel uses the same $I_{\max}$, a stronger string cloud does not simply brighten the entire image; instead, it redistributes the radiative flux among the image branches. The peak and screen-integrated intensities therefore need not increase monotonically with $\mathcal A$.

\begin{figure*}[htp!]
    \centering
    \includegraphics[width=0.85\linewidth]{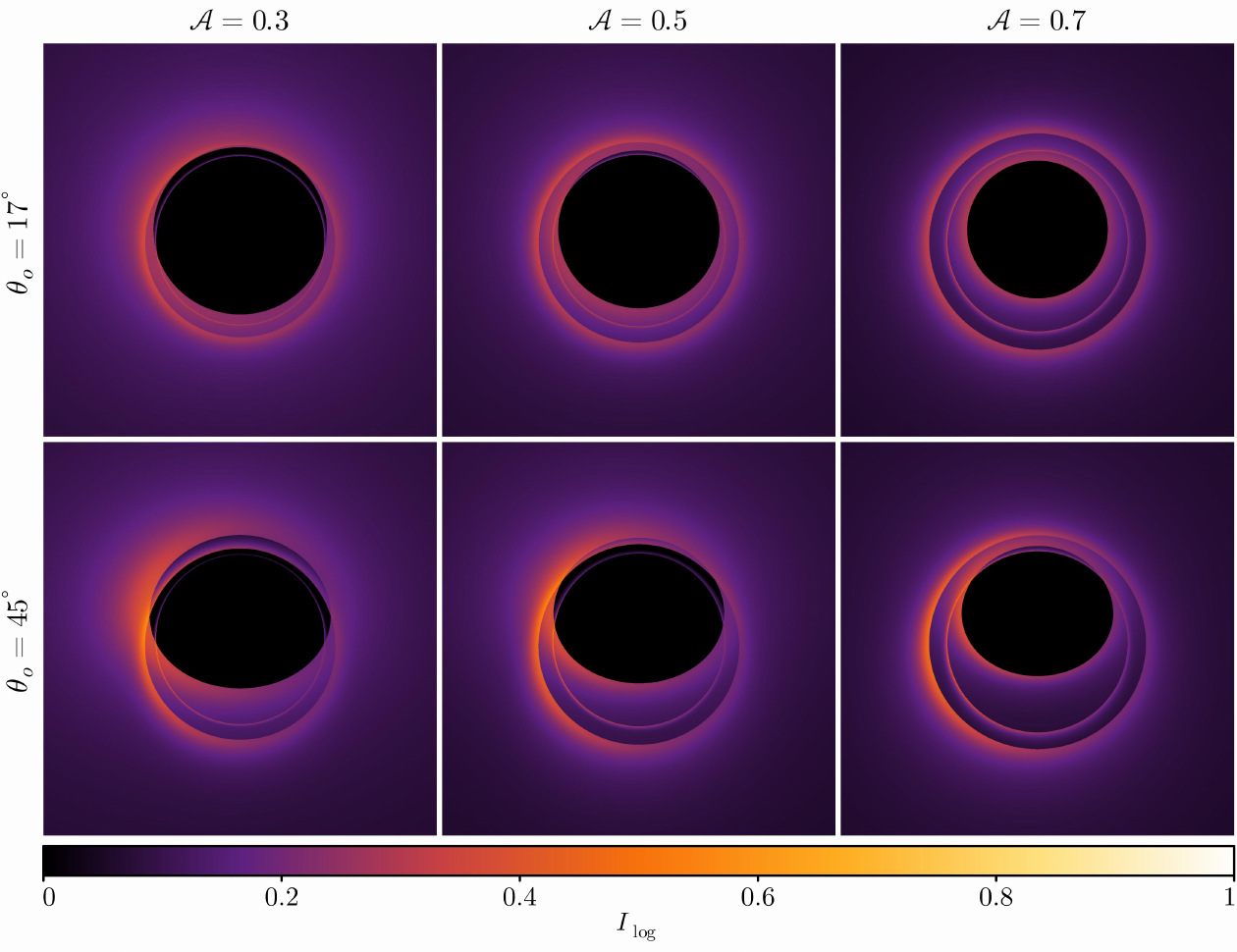}
    \caption{Inclination dependence of the logarithmically mapped thin-disk image at fixed $a_0/r_+=2.0$. The columns show $\mathcal A=0.3$, $0.5$, and $0.7$, whereas the upper and lower rows show $\theta_o=17^\circ$ and $45^\circ$, respectively. In every panel, $M=1$, $r_0/M=0.20$, $r_o=200M$, and the screen half-width is $2.4b_{\rm ph}$. All six panels use the same GLM emissivity and the common normalization of the 18 configurations in the primary scan. For these parameters, Table~\ref{tab:optical_scales} shows that the throat lies outside the outer photon sphere.}
    \label{fig:wormhole_thin_disk_inclinations}
\end{figure*}

Figure~\ref{fig:wormhole_thin_disk_inclinations} isolates the effect of viewing inclination at $a_0/r_+=2.0$ by presenting the three values of $\mathcal A$ at $\theta_o=17^\circ$ and $45^\circ$. Together with the $\theta_o=80^\circ$ panels in Fig.~\ref{fig:wormhole_thin_disk_images}, it provides a comparison across all three inclinations. All images use the same normalization, so no panel-by-panel rescaling is hidden in the comparison.

At $17^\circ$, the disk is viewed almost face-on, so both its direct and lensed images appear as nearly concentric annuli and the azimuthal Doppler asymmetry is weak. At $45^\circ$, projection gradually separates the upper and lower arcs, and the approaching edge of the disk becomes noticeably brighter. Comparison with the $80^\circ$ panels in Fig.~\ref{fig:wormhole_thin_disk_images} shows that the direct image then contracts into a narrow band, the approaching-side enhancement is strongest, and the higher-order rings are most clearly separated from the primary image. The raw data show the same trend: for the three values of $\mathcal A$ at $a_0/r_+=2.0$, the peak $I_o$ increases from $0.49$--$0.71$ at $17^\circ$ to $1.32$--$1.58$ at $45^\circ$ and further to $3.53$--$3.78$ at $80^\circ$. Thus, although the spacetime is spherically symmetric, the accretion disk is only axisymmetric, and the viewing inclination changes both the image morphology and observed intensity through projection and Doppler weighting. At each inclination, increasing $\mathcal A$ reveals additional annular substructure.

The calculations at $a_0/r_+=1.2$ show the same qualitative inclination trend: a nearly face-on view is dominated by approximately concentric rings, while increasing the inclination strengthens the left--right brightness asymmetry and stretches the higher-order images into more distinct arcs. The principal difference is not the inclination dependence itself. Because the smaller throat retains the outer photon sphere, it supports a denser set of narrow higher-order rings at each inclination. This throat-radius dependence is already displayed directly by the two rows of Fig.~\ref{fig:wormhole_thin_disk_images} at $\theta_o=80^\circ$ and by the celestial-sphere images in Fig.~\ref{fig:wormhole_lensing_images}.

\begin{figure*}[htp!]
    \centering
    \includegraphics[width=0.85\linewidth]{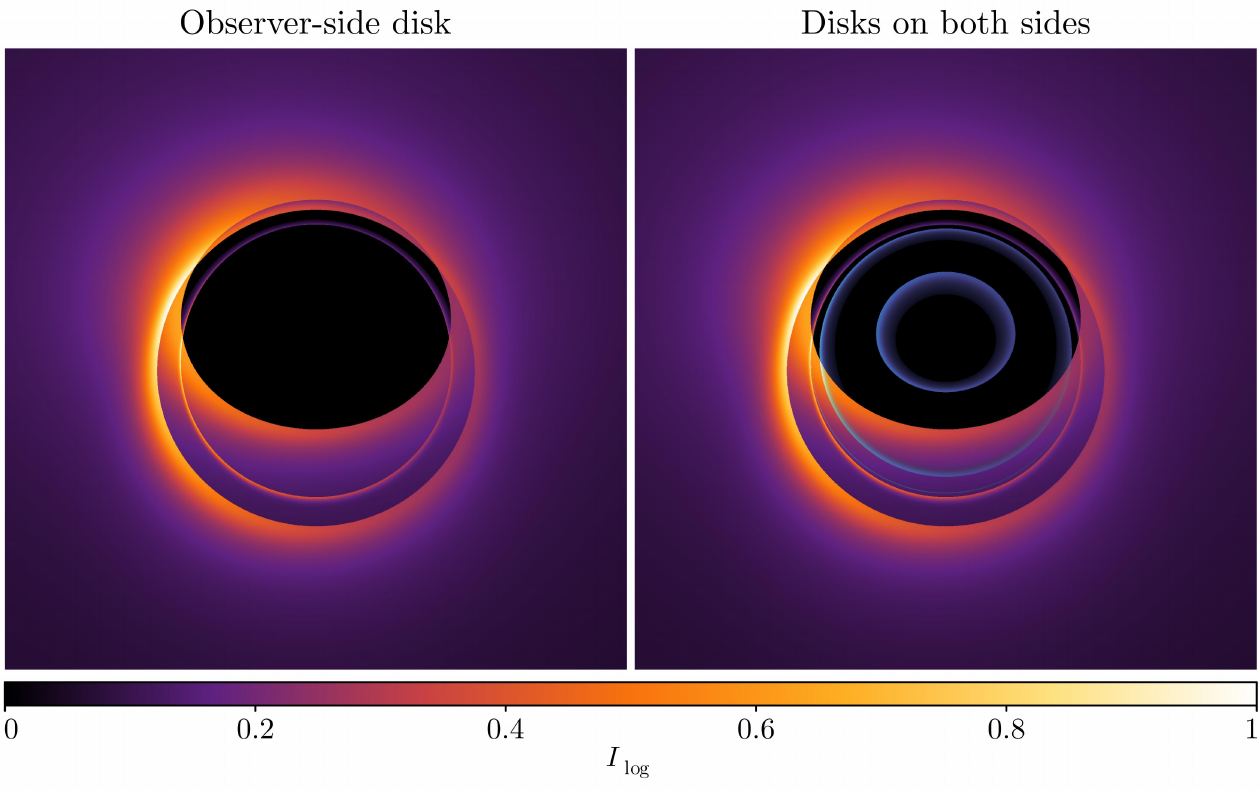}
    \caption{Comparison of an observer-side disk (left) with identical disks in both exterior regions (right), for $(\mathcal A,a_0/r_+,\theta_o)=(0.5,2.0,45^\circ)$ and $M=1$, $r_0/M=0.20$, and $r_o=200M$. Both panels use the same pairwise intensity normalization. Warm colors denote observer-side emission; in the right panel, blue denotes the contribution from the disk in the opposite universe, and mixed colors occur where the two contributions overlap. The bottom scale gives the common nonlinear intensity coordinate $I_{\log}$, whereas hue also identifies the side from which the radiation originates.}
    \label{fig:wormhole_two_sided_disks}
\end{figure*}

Finally, Fig.~\ref{fig:wormhole_two_sided_disks} distinguishes higher-order images of the disk on the observer's side from radiation that genuinely originates in the second exterior. The left panel contains only the observer-side disk, as in the primary scan. The right panel adds an identical equatorial disk in the other universe without changing the geometry, observer, or observer-side emission. Both panels use the same pairwise value of $I_{\max}$ and can therefore be compared quantitatively within the figure; however, their color scale should not be compared directly with the common 18-configuration scale used in Figs.~\ref{fig:wormhole_thin_disk_images} and \ref{fig:wormhole_thin_disk_inclinations}.

After the second disk is added, all warm observer-side structures remain unchanged, but a blue ring appears inside the previously dark central region, together with thinner blue or cyan counterimages near the same-side image branches. These structures are produced by null geodesics that cross the throat, intersect the far-side disk, and then reach the observer. On the adopted $2000^2$-pixel screen, the far-side disk contributes to approximately $5.1\%$ of the pixels and increases the mean raw intensity over the screen by about $8.0\%$. This explicit encoding of the emission side shows that the innermost rings are not all higher-order images of the observer-side disk; some are genuine optical images of matter in the second asymptotic region. The result also confirms that, in a one-sided emission model, a dark region alone does not demonstrate that the corresponding null geodesics are captured.

\section{Conclusion}\label{sec:conclusion}

In this work, we constructed a symmetric thin-shell wormhole by joining two copies of the positive-lapse exterior of the rationally regularized Letelier--Alencar string-cloud black hole. The resulting manifold has two exterior ends connected through a timelike minimum-area surface located outside the outer horizon. Although the seed spacetime is nonsingular and possesses a finite-curvature anti--de Sitter core, the junction excises the core and the horizons from both copies. Regularity of the bulk therefore provides a well-defined seed geometry but does not, by itself, eliminate the exotic surface matter required by the symmetric flare-out construction.

The Darmois--Israel junction conditions show analytically that every admissible static throat has negative surface energy density. Consequently, the intrinsic weak and dominant energy conditions are violated throughout the static exterior. The null energy condition restricted to directions tangent to the shell is satisfied only in an inner radial interval and becomes violated after a single transition. Since the tangential pressure remains positive in the sampled domain, the standard intrinsic $2+1$-dimensional strong energy condition has that same inner domain. The auxiliary combination $\sigma_0+2p_0$ remains positive, but it is not by itself the intrinsic strong condition. Moreover, the complete four-dimensional distributional null energy condition remains violated because radial null directions probe the negative surface density. Near a simple outer horizon, the surface density tends to zero from below while the tangential pressure becomes dominant, so this limit cannot be interpreted as the disappearance of the shell or as a regular matter-free configuration.

The radial-vacuum relation obeyed by the regularized string-cloud source removes the net momentum flux across the moving shell. This reduces the conservation law to its standard intrinsic form and allows the throat dynamics to be written in terms of an effective potential. We obtained a general local-stability criterion applicable to surface equations of state with both density dependence and explicit radial dependence. For every model, one equation-of-state parameter was fixed by the static junction condition and then held constant during the perturbation. This distinction is essential: varying the equilibrium radius generates a family of separately calibrated static shells rather than the dynamical trajectory of a single wormhole.

The stability analysis demonstrates that the shell response depends strongly on its phenomenological matter model. No locally stable linear barotropic configuration was identified in the explored exterior, although the instability becomes progressively weaker at large radii. Introducing explicit radial dependence through the variable phantom law produces stable outer branches, with larger values of its radial exponent moving the stability transition closer to the horizon and enlarging the stable portion of the sampled family. The variable Chaplygin model displays the opposite near-horizon behavior: all configurations examined begin on a locally stable branch, while the case with exponent below two eventually becomes weakly unstable farther outward. For exponents equal to or greater than two, the potential curvature remains positive throughout the plotted interval, although it approaches marginality at large radius.

For the generalized cosmic Chaplygin gas, the selected continuous real branch consistently evolves from an unstable inner regime to a stable outer branch. The modified cosmic Chaplygin gas exhibits a stronger dependence on the cosmic parameter. For $\omega=-2$, systematic outer stability is absent from almost all configurations examined, apart from a narrow positive segment in one selected case. For $\omega=-10$, all sampled branches cross from negative to positive potential curvature and retain an outer stable regime. These differences cannot be attributed solely to the explicit linear modification, because the static calibration simultaneously changes the nonlinear Chaplygin parameter. The string-cloud strength likewise does not act as a uniformly stabilizing parameter: in several models it reduces the magnitudes of both stable and unstable branches and shifts the configurations toward marginality.

The related string-cloud thin-shell results provide two complementary checks on this interpretation. The exact $r_0=0$ junction and general radial-stability expressions reproduce the singular Letelier formulas of Richarte and Simeone after the notation and equation-of-state slopes are matched \cite{RicharteSimeone2008}. Across the Bardeen--AdS and Hayward--Letelier--AdS comparisons, the same calibrated variable closures produce seed-dependent exponent trends \cite{WaseemEtAl2025StringFluid,HouedokouEtAl2026HaywardLetelier}. Together with the analytic controls above, these comparisons establish that the observed stability branches arise from the coupled choice of exterior lapse and shell response, rather than from regularity or string matter in isolation.

Comparison with the exact unregularized exterior shows that several principal non-optical sign structures are not created by the regular core. Negative surface density and the inner-to-outer change of the tangential null-energy combination persist analytically, as do the instability of the barotropic shell and the characteristic stable branches of the variable phantom and variable Chaplygin models. For the cosmic Chaplygin closures, however, a real and simple calibration branch does not guarantee preservation of the stability sign: the MCCG example discussed above crosses $V''_{\rm M}=0$ during continuation to $r_0=0$ without losing the selected root. The regulator modifies the quantitative location and strength of the verified features through the retained exterior lapse, but the fixed-parameter comparison does not isolate the regular core from concomitant changes in physical radius, horizon scale, and asymptotic mass.

The optical analysis provides a complementary signature of both the string cloud and the junction. Two separately color-coded celestial spheres distinguish rays reaching the celestial sphere in the observer's exterior from through-throat rays reaching the celestial sphere in the second exterior, while a point source aligned with the observer and wormhole center produces multiple Einstein rings. Moving the throat reorganizes mainly the inner, shell-probing branches while leaving the exterior direct image nearly unchanged. In particular, placing the throat outside the seed photon sphere removes its photon-ring sequence but not the broader ringlike images produced by strong deflection and through-throat propagation. Increasing $\mathcal A$ changes the spacing and visibility of the higher-order domains. Thin-disk images show an additional strong inclination dependence: an almost face-on system is dominated by concentric annuli, whereas a nearly edge-on disk develops a narrow direct band, strong Doppler asymmetry, and well-separated lensed arcs. Greater image multiplicity does not imply uniform brightening under the common intensity normalization. When identical disks are placed in both universes, through-throat radiation produces distinct central and inner counterimages that are absent for a one-sided disk, thereby separating genuine far-side emission from same-side higher-order images. A dark central region in the one-sided model therefore depends on the assumed source distribution and disk truncation, as well as on the null-geodesic structure, and should not automatically be identified with a black-hole capture shadow.

The conclusions are restricted to symmetric spherical junctions, phenomenological surface equations of state, selected real Chaplygin branches, and linear radial perturbations over finite parameter ranges. The optical results additionally rely on prescribed celestial-sphere and thin-disk source models, transparent throat crossing, a phenomenological emissivity profile, and the neglect of absorption, scattering, disk thickness, and disk backreaction. They do not establish nonlinear, nonspherical, or global stability, and the horizon limit itself is excluded from the timelike construction. Nevertheless, the symmetric construction developed here provides a natural benchmark against which the effects introduced by adopting inequivalent geometries on the two sides of the throat can be systematically isolated. Natural extensions are therefore to examine asymmetric junctions, nonlinear and nonspherical perturbations, and alternative real branches supported by a more explicit microphysical model of the shell, as well as radiative-transfer models incorporating finite disk thickness, absorption, scattering, and dynamically consistent accretion. A future mass-matched analysis at a common physical radius could sharpen the quantitative separation between regulator and asymptotic-mass effects beyond the fixed-parameter analytic control used here.

\begin{acknowledgments}
EO would like to thank Fundação Cearense de Apoio ao Desenvolvimento Científico e Tecnológico (FUNCAP), through grant BP6-0241-00335.01.00/25. This work was also supported by NSFC under Grant No.~12575048. M.G. also acknowledges support from the BNU Tang Scholar Program.
\end{acknowledgments}

\bibliography{references}

@article{MorrisThorne1988,
  author = {Morris, Michael S. and Thorne, Kip S.},
  title = {Wormholes in spacetime and their use for interstellar travel: A tool for teaching general relativity},
  journal = {American Journal of Physics},
  volume = {56},
  number = {5},
  pages = {395--412},
  year = {1988},
  doi = {10.1119/1.15620},
  url = {https://doi.org/10.1119/1.15620}
}

@article{MorrisThorneYurtsever1988,
  author = {Morris, Michael S. and Thorne, Kip S. and Yurtsever, Ulvi},
  title = {Wormholes, time machines, and the weak energy condition},
  journal = {Physical Review Letters},
  volume = {61},
  number = {13},
  pages = {1446--1449},
  year = {1988},
  doi = {10.1103/PhysRevLett.61.1446},
  url = {https://doi.org/10.1103/PhysRevLett.61.1446}
}

@article{FriedmanSchleichWitt1993,
  author = {Friedman, John L. and Schleich, Kristin and Witt, Donald M.},
  title = {Topological censorship},
  journal = {Physical Review Letters},
  volume = {71},
  number = {10},
  pages = {1486--1489},
  year = {1993},
  doi = {10.1103/PhysRevLett.71.1486},
  url = {https://doi.org/10.1103/PhysRevLett.71.1486}
}

@article{FordRoman1996,
  author = {Ford, L. H. and Roman, Thomas A.},
  title = {Quantum field theory constrains traversable wormhole geometries},
  journal = {Physical Review D},
  volume = {53},
  number = {10},
  pages = {5496--5507},
  year = {1996},
  doi = {10.1103/PhysRevD.53.5496},
  url = {https://doi.org/10.1103/PhysRevD.53.5496}
}

@article{FewsterRoman2005,
  author = {Fewster, Christopher J. and Roman, Thomas A.},
  title = {On wormholes with arbitrarily small quantities of exotic matter},
  journal = {Physical Review D},
  volume = {72},
  number = {4},
  pages = {044023},
  year = {2005},
  doi = {10.1103/PhysRevD.72.044023},
  url = {https://doi.org/10.1103/PhysRevD.72.044023}
}

@article{Letelier1979,
  author = {Letelier, Patricio S.},
  title = {Clouds of strings in general relativity},
  journal = {Physical Review D},
  volume = {20},
  number = {6},
  pages = {1294--1302},
  year = {1979},
  doi = {10.1103/PhysRevD.20.1294},
  url = {https://doi.org/10.1103/PhysRevD.20.1294}
}

@article{Alencar2025,
  author = {Alencar, G. and Landim, R. R. and Costa Filho, R. N.},
  title = {A new cloud of strings},
  journal = {Physics of the Dark Universe},
  volume = {49},
  pages = {102031},
  year = {2025},
  doi = {10.1016/j.dark.2025.102031},
  url = {https://doi.org/10.1016/j.dark.2025.102031}
}

@article{Santos2025StringFluids,
  author = {Santos, Luis C. N.},
  title = {Revisiting black holes surrounded by cloud and fluid of strings in general relativity},
  journal = {Physical Review D},
  volume = {111},
  number = {6},
  pages = {064032},
  year = {2025},
  doi = {10.1103/PhysRevD.111.064032},
  url = {https://doi.org/10.1103/PhysRevD.111.064032}
}

@article{Muniz2026,
  author = {Muniz, Celio R. and Rebouças, Jonathan Alves and de Oliveira, Leonardo Tavares and Sampaio, Francisco Tiago Barboza and Lustosa, Francisco Bento},
  title = {Regularized black hole solution from a new string cloud source},
  journal = {Physics of the Dark Universe},
  volume = {52},
  pages = {102272},
  year = {2026},
  doi = {10.1016/j.dark.2026.102272},
  url = {https://doi.org/10.1016/j.dark.2026.102272}
}

@article{Hayward2006,
  author = {Hayward, Sean A.},
  title = {Formation and evaporation of nonsingular black holes},
  journal = {Physical Review Letters},
  volume = {96},
  number = {3},
  pages = {031103},
  year = {2006},
  doi = {10.1103/PhysRevLett.96.031103},
  url = {https://doi.org/10.1103/PhysRevLett.96.031103}
}

@article{Frolov2016,
  author = {Frolov, Valeri P.},
  title = {Notes on nonsingular models of black holes},
  journal = {Physical Review D},
  volume = {94},
  number = {10},
  pages = {104056},
  year = {2016},
  doi = {10.1103/PhysRevD.94.104056},
  url = {https://doi.org/10.1103/PhysRevD.94.104056}
}

@article{Zaslavskii2010,
  author = {Zaslavskii, O. B.},
  title = {Regular black holes and energy conditions},
  journal = {Physics Letters B},
  volume = {688},
  number = {4-5},
  pages = {278--280},
  year = {2010},
  doi = {10.1016/j.physletb.2010.04.031},
  url = {https://doi.org/10.1016/j.physletb.2010.04.031}
}

@article{BalartVagenas2014,
  author = {Balart, Leonardo and Vagenas, Elias C.},
  title = {Regular black hole metrics and the weak energy condition},
  journal = {Physics Letters B},
  volume = {730},
  pages = {14--17},
  year = {2014},
  doi = {10.1016/j.physletb.2014.01.024},
  url = {https://doi.org/10.1016/j.physletb.2014.01.024}
}

@article{Israel1966,
  author = {Israel, Werner},
  title = {Singular hypersurfaces and thin shells in general relativity},
  journal = {Il Nuovo Cimento B},
  volume = {44},
  number = {1},
  pages = {1--14},
  year = {1966},
  doi = {10.1007/BF02710419},
  url = {https://doi.org/10.1007/BF02710419}
}

@article{Visser1989,
  author = {Visser, Matt},
  title = {Traversable wormholes from surgically modified Schwarzschild spacetimes},
  journal = {Nuclear Physics B},
  volume = {328},
  number = {1},
  pages = {203--212},
  year = {1989},
  doi = {10.1016/0550-3213(89)90100-4},
  url = {https://doi.org/10.1016/0550-3213(89)90100-4}
}

@article{PoissonVisser1995,
  author = {Poisson, Eric and Visser, Matt},
  title = {Thin-shell wormholes: Linearization stability},
  journal = {Physical Review D},
  volume = {52},
  number = {12},
  pages = {7318--7321},
  year = {1995},
  doi = {10.1103/PhysRevD.52.7318},
  url = {https://doi.org/10.1103/PhysRevD.52.7318}
}

@article{Eiroa2008SphericalStability,
  author = {Eiroa, Ernesto F.},
  title = {Stability of thin-shell wormholes with spherical symmetry},
  journal = {Physical Review D},
  volume = {78},
  number = {2},
  pages = {024018},
  year = {2008},
  doi = {10.1103/PhysRevD.78.024018},
  url = {https://doi.org/10.1103/PhysRevD.78.024018}
}

@article{GarciaLoboVisser2012,
  author = {García, Nadiezhda Montelongo and Lobo, Francisco S. N. and Visser, Matt},
  title = {Generic spherically symmetric dynamic thin-shell traversable wormholes in standard general relativity},
  journal = {Physical Review D},
  volume = {86},
  number = {4},
  pages = {044026},
  year = {2012},
  doi = {10.1103/PhysRevD.86.044026},
  url = {https://doi.org/10.1103/PhysRevD.86.044026}
}

@article{LoboCrawford2004,
  author = {Lobo, Francisco S. N. and Crawford, Paulo},
  title = {Linearized stability analysis of thin-shell wormholes with a cosmological constant},
  journal = {Classical and Quantum Gravity},
  volume = {21},
  number = {2},
  pages = {391--404},
  year = {2004},
  doi = {10.1088/0264-9381/21/2/004},
  url = {https://doi.org/10.1088/0264-9381/21/2/004}
}

@article{EiroaSimeone2007,
  author = {Eiroa, Ernesto F. and Simeone, Claudio},
  title = {Stability of Chaplygin gas thin-shell wormholes},
  journal = {Physical Review D},
  volume = {76},
  number = {2},
  pages = {024021},
  year = {2007},
  doi = {10.1103/PhysRevD.76.024021},
  url = {https://doi.org/10.1103/PhysRevD.76.024021},
  eprint = {0704.1136},
  archivePrefix = {arXiv},
  primaryClass = {gr-qc}
}

@article{Eiroa2009GCG,
  author = {Eiroa, Ernesto F.},
  title = {Thin-shell wormholes with a generalized Chaplygin gas},
  journal = {Physical Review D},
  volume = {80},
  number = {4},
  pages = {044033},
  year = {2009},
  doi = {10.1103/PhysRevD.80.044033},
  url = {https://doi.org/10.1103/PhysRevD.80.044033}
}

@article{SharifAzam2013GCCG,
  author = {Sharif, M. and Azam, M.},
  title = {Reissner--Nordstr\"om thin-shell wormholes with generalized cosmic Chaplygin gas},
  journal = {The European Physical Journal C},
  volume = {73},
  pages = {2554},
  year = {2013},
  doi = {10.1140/epjc/s10052-013-2554-z},
  url = {https://doi.org/10.1140/epjc/s10052-013-2554-z}
}

@article{SharifMumtaz2014,
  author = {Sharif, M. and Mumtaz, Saadia},
  title = {Schwarzschild--de Sitter and anti--de Sitter thin-shell wormholes and their stability},
  journal = {Advances in High Energy Physics},
  volume = {2014},
  pages = {639759},
  year = {2014},
  doi = {10.1155/2014/639759},
  url = {https://doi.org/10.1155/2014/639759}
}

@article{HalilsoyOvgunMazharimousavi2014,
  author = {Halilsoy, M. and Övgün, A. and Mazharimousavi, S. H.},
  title = {Thin-shell wormholes from the regular Hayward black hole},
  journal = {The European Physical Journal C},
  volume = {74},
  pages = {2796},
  year = {2014},
  doi = {10.1140/epjc/s10052-014-2796-4},
  url = {https://doi.org/10.1140/epjc/s10052-014-2796-4}
}

@article{SharifMumtaz2016Hayward,
  author = {Sharif, M. and Mumtaz, Saadia},
  title = {Stability of the regular Hayward thin-shell wormholes},
  journal = {Advances in High Energy Physics},
  volume = {2016},
  pages = {2868750},
  year = {2016},
  doi = {10.1155/2016/2868750},
  url = {https://doi.org/10.1155/2016/2868750}
}

@article{Kuhfittig2010,
  author = {Kuhfittig, Peter K. F.},
  title = {The stability of thin-shell wormholes with a phantom-like equation of state},
  journal = {Acta Physica Polonica B},
  volume = {41},
  pages = {2017--2029},
  year = {2010},
  url = {https://www.actaphys.uj.edu.pl/R/41/9/2017/pdf}
}

@article{Varela2015,
  author = {Varela, Victor},
  title = {Note on linearized stability of Schwarzschild thin-shell wormholes with variable equations of state},
  journal = {Physical Review D},
  volume = {92},
  number = {4},
  pages = {044002},
  year = {2015},
  doi = {10.1103/PhysRevD.92.044002},
  url = {https://doi.org/10.1103/PhysRevD.92.044002}
}

@article{Javed2024,
  author = {Javed, Faisal and Waseem, Arfa and Fatima, Ghulam and Almutairi, Bander},
  title = {Stability of thin-shell wormholes via polymer black hole in loop quantum gravity},
  journal = {Physics of the Dark Universe},
  volume = {46},
  pages = {101605},
  year = {2024},
  doi = {10.1016/j.dark.2024.101605},
  url = {https://doi.org/10.1016/j.dark.2024.101605}
}

@article{GonzalezDiaz2003,
  author = {González-Díaz, Pedro F.},
  title = {You need not be afraid of phantom energy},
  journal = {Physical Review D},
  volume = {68},
  number = {2},
  pages = {021303},
  year = {2003},
  doi = {10.1103/PhysRevD.68.021303},
  url = {https://doi.org/10.1103/PhysRevD.68.021303}
}

@article{SadeghiFarahani2013,
  author = {Sadeghi, Jafar and Farahani, Hoda},
  title = {Interaction between viscous varying modified cosmic Chaplygin gas and tachyonic fluid},
  journal = {Astrophysics and Space Science},
  volume = {347},
  pages = {209--219},
  year = {2013},
  doi = {10.1007/s10509-013-1507-1},
  url = {https://doi.org/10.1007/s10509-013-1507-1}
}

@article{HochbergVisser1998,
  author = {Hochberg, David and Visser, Matt},
  title = {The null energy condition in dynamic wormholes},
  journal = {Physical Review Letters},
  volume = {81},
  number = {4},
  pages = {746--749},
  year = {1998},
  doi = {10.1103/PhysRevLett.81.746},
  url = {https://doi.org/10.1103/PhysRevLett.81.746}
}

@article{VisserKarDadhich2003,
  author = {Visser, Matt and Kar, Sayan and Dadhich, Naresh},
  title = {Traversable wormholes with arbitrarily small energy condition violations},
  journal = {Physical Review Letters},
  volume = {90},
  number = {20},
  pages = {201102},
  year = {2003},
  doi = {10.1103/PhysRevLett.90.201102},
  url = {https://doi.org/10.1103/PhysRevLett.90.201102}
}

@article{EiroaEtAl2026ScalarTensor,
  author = {Eiroa, Ernesto F. and Figueroa-Aguirre, Griselda and Karanasou, Vasiliki},
  title = {Stability of spherical thin-shell wormholes in scalar-tensor theories},
  journal = {The European Physical Journal Plus},
  volume = {141},
  number = {2},
  pages = {134},
  year = {2026},
  doi = {10.1140/epjp/s13360-026-07350-z},
  url = {https://doi.org/10.1140/epjp/s13360-026-07350-z}
}

@article{EiroaEtAl2024ChargedThermo,
  author = {Eiroa, Ernesto F. and Figueroa-Aguirre, Griselda and Pe{\~n}afiel, Miguel L. and Perez Bergliaffa, Santiago Esteban},
  title = {Dynamical and thermodynamical stability of a charged thin-shell wormhole},
  journal = {The European Physical Journal C},
  volume = {84},
  number = {11},
  pages = {1160},
  year = {2024},
  doi = {10.1140/epjc/s10052-024-13465-3},
  url = {https://doi.org/10.1140/epjc/s10052-024-13465-3}
}

@article{JavedEtAl2023QuantumCorrected,
  author = {Javed, Faisal and Waseem, Arfa and Almutairi, Bander},
  title = {Quantum corrected charged thin-shell wormholes surrounded by quintessence},
  journal = {The European Physical Journal C},
  volume = {83},
  number = {9},
  pages = {811},
  year = {2023},
  doi = {10.1140/epjc/s10052-023-11990-1},
  url = {https://doi.org/10.1140/epjc/s10052-023-11990-1}
}

@article{KokubuHarada2020,
  author = {Kokubu, Takafumi and Harada, Tomohiro},
  title = {Thin-Shell Wormholes in Einstein and Einstein--Gauss--Bonnet Theories of Gravity},
  journal = {Universe},
  volume = {6},
  number = {11},
  pages = {197},
  year = {2020},
  doi = {10.3390/universe6110197},
  url = {https://doi.org/10.3390/universe6110197}
}

@article{ForghaniEtAl2020EnergyConditions,
  author = {Forghani, S. Danial and Mazharimousavi, S. Habib and Halilsoy, Mustafa},
  title = {Thin-shell wormhole satisfying energy conditions},
  journal = {Physics Letters B},
  volume = {804},
  pages = {135374},
  year = {2020},
  doi = {10.1016/j.physletb.2020.135374},
  url = {https://doi.org/10.1016/j.physletb.2020.135374}
}

@article{LoboSimpsonVisser2020,
  author = {Lobo, Francisco S. N. and Simpson, Alex and Visser, Matt},
  title = {Dynamic thin-shell black-bounce traversable wormholes},
  journal = {Physical Review D},
  volume = {101},
  number = {12},
  pages = {124035},
  year = {2020},
  doi = {10.1103/PhysRevD.101.124035},
  url = {https://doi.org/10.1103/PhysRevD.101.124035}
}

@article{ForghaniEtAl2019Thermodynamic,
  author = {Forghani, S. Danial and Mazharimousavi, S. Habib and Halilsoy, Mustafa},
  title = {Thermodynamic stability of a Schwarzschild thin-shell wormhole},
  journal = {International Journal of Modern Physics D},
  volume = {28},
  number = {11},
  pages = {1950142},
  year = {2019},
  doi = {10.1142/S0218271819501426},
  url = {https://doi.org/10.1142/S0218271819501426}
}

@article{MazharimousaviHalilsoy2014FlareOut,
  author = {Mazharimousavi, S. Habib and Halilsoy, M.},
  title = {Flare-out conditions in static thin-shell wormholes},
  journal = {Physical Review D},
  volume = {90},
  number = {8},
  pages = {087501},
  year = {2014},
  doi = {10.1103/PhysRevD.90.087501},
  url = {https://doi.org/10.1103/PhysRevD.90.087501}
}

@article{EiroaFigueroaAguirre2012BornInfeld,
  author = {Eiroa, Ernesto F. and Figueroa-Aguirre, Griselda},
  title = {Thin-shell wormholes with a generalized Chaplygin gas in Einstein--Born--Infeld theory},
  journal = {The European Physical Journal C},
  volume = {72},
  number = {11},
  pages = {2240},
  year = {2012},
  doi = {10.1140/epjc/s10052-012-2240-6},
  url = {https://doi.org/10.1140/epjc/s10052-012-2240-6}
}

@article{DiasLemos2010,
  author = {Dias, Gon{\c{c}}alo A. S. and Lemos, Jos{\'e} P. S.},
  title = {Thin-shell wormholes in {$d$}-dimensional general relativity: Solutions, properties, and stability},
  journal = {Physical Review D},
  volume = {82},
  number = {8},
  pages = {084023},
  year = {2010},
  doi = {10.1103/PhysRevD.82.084023},
  url = {https://doi.org/10.1103/PhysRevD.82.084023}
}

@article{EiroaRomero2004Charged,
  author = {Eiroa, Ernesto F. and Romero, Gustavo E.},
  title = {Linearized Stability of Charged Thin-Shell Wormholes},
  journal = {General Relativity and Gravitation},
  volume = {36},
  number = {4},
  pages = {651--659},
  year = {2004},
  doi = {10.1023/B:GERG.0000016916.79221.24},
  url = {https://doi.org/10.1023/B:GERG.0000016916.79221.24}
}

@article{Sushkov2005Phantom,
  author = {Sushkov, S. V.},
  title = {Wormholes supported by a phantom energy},
  journal = {Physical Review D},
  volume = {71},
  number = {4},
  pages = {043520},
  year = {2005},
  doi = {10.1103/PhysRevD.71.043520},
  url = {https://doi.org/10.1103/PhysRevD.71.043520},
  eprint = {gr-qc/0502084},
  archivePrefix = {arXiv}
}

@article{Dymnikova2004RegularCharged,
  author = {Dymnikova, Irina},
  title = {Regular electrically charged vacuum structures with de Sitter centre in nonlinear electrodynamics coupled to general relativity},
  journal = {Classical and Quantum Gravity},
  volume = {21},
  number = {18},
  pages = {4417--4428},
  year = {2004},
  doi = {10.1088/0264-9381/21/18/009},
  url = {https://doi.org/10.1088/0264-9381/21/18/009}
}

@article{Bronnikov2001RegularMagnetic,
  author = {Bronnikov, K. A.},
  title = {Regular magnetic black holes and monopoles from nonlinear electrodynamics},
  journal = {Physical Review D},
  volume = {63},
  number = {4},
  pages = {044005},
  year = {2001},
  doi = {10.1103/PhysRevD.63.044005},
  url = {https://doi.org/10.1103/PhysRevD.63.044005}
}

@article{AyonBeatoGarcia1998,
  author = {Ay{\'o}n-Beato, Eloy and Garc{\'i}a, Alberto},
  title = {Regular Black Hole in General Relativity Coupled to Nonlinear Electrodynamics},
  journal = {Physical Review Letters},
  volume = {80},
  number = {23},
  pages = {5056--5059},
  year = {1998},
  doi = {10.1103/PhysRevLett.80.5056},
  url = {https://doi.org/10.1103/PhysRevLett.80.5056}
}

@article{AyonBeatoGarcia2005,
  author = {Ay{\'o}n-Beato, Eloy and Garc{\'i}a, Alberto},
  title = {Four-parametric regular black hole solution},
  journal = {General Relativity and Gravitation},
  volume = {37},
  number = {4},
  pages = {635--641},
  year = {2005},
  doi = {10.1007/s10714-005-0050-y},
  url = {https://doi.org/10.1007/s10714-005-0050-y}
}

@article{ReboucasEtAl2026QuantumImproved,
  author = {Rebou{\c{c}}as, Jonathan A. and Muniz, Celio R. and Lustosa, Francisco Bento and Otoniel, Edson},
  title = {Quantum-improved wormholes in the Dekel--Zhao dark matter halo},
  journal = {The European Physical Journal C},
  volume = {86},
  number = {4},
  pages = {407},
  year = {2026},
  doi = {10.1140/epjc/s10052-026-15676-2},
  url = {https://doi.org/10.1140/epjc/s10052-026-15676-2}
}

@article{Tsukamoto2021DoubleShadows,
  author = {Tsukamoto, Naoki},
  title = {Linearization stability of reflection-asymmetric thin-shell wormholes with double shadows},
  journal = {Physical Review D},
  volume = {103},
  number = {6},
  pages = {064031},
  year = {2021},
  doi = {10.1103/PhysRevD.103.064031},
  url = {https://doi.org/10.1103/PhysRevD.103.064031}
}

@article{TangphatiEtAl2020dRGT,
  author = {Tangphati, Takol and Chatrabhuti, Auttakit and Samart, Daris and Channuie, Phongpichit},
  title = {Thin-shell wormholes in de Rham--Gabadadze--Tolley massive gravity},
  journal = {The European Physical Journal C},
  volume = {80},
  number = {8},
  pages = {722},
  year = {2020},
  doi = {10.1140/epjc/s10052-020-8294-y},
  url = {https://doi.org/10.1140/epjc/s10052-020-8294-y}
}

@article{IshibashiOhtaYamaguchi2021,
  author = {Ishibashi, Akihiro and Ohta, Nobuyoshi and Yamaguchi, Daiki},
  title = {Quantum improved charged black holes},
  journal = {Physical Review D},
  volume = {104},
  number = {6},
  pages = {066016},
  year = {2021},
  doi = {10.1103/PhysRevD.104.066016},
  url = {https://doi.org/10.1103/PhysRevD.104.066016}
}

@article{RadhakrishnanEtAl2024Review,
  author = {Radhakrishnan, Ramesh and Brown, Patrick and Matulevich, Jacob and Davis, Eric and Mirfendereski, Delaram and Cleaver, Gerald},
  title = {A Review of Stable, Traversable Wormholes in {$f(R)$} Gravity Theories},
  journal = {Symmetry},
  volume = {16},
  number = {8},
  pages = {1007},
  year = {2024},
  doi = {10.3390/sym16081007},
  url = {https://doi.org/10.3390/sym16081007}
}

@article{TsukamotoKokubu2024PhotonOrbits,
  author = {Tsukamoto, Naoki and Kokubu, Takafumi},
  title = {Relation between circular photon orbits and the stability of wormholes with the thin shell of a barotropic fluid},
  journal = {Physical Review D},
  volume = {109},
  number = {2},
  pages = {024047},
  year = {2024},
  doi = {10.1103/PhysRevD.109.024047},
  url = {https://doi.org/10.1103/PhysRevD.109.024047},
  eprint = {2305.07256},
  archivePrefix = {arXiv},
  primaryClass = {gr-qc}
}

@article{Mazharimousavi2024InnermostPhotonSphere,
  author = {Mazharimousavi, S. Habib},
  title = {Generic spherically symmetric thin-shell wormholes with equilibrium throat at the innermost photonsphere are unstable},
  journal = {The European Physical Journal C},
  volume = {84},
  pages = {303},
  year = {2024},
  doi = {10.1140/epjc/s10052-024-12678-w},
  url = {https://doi.org/10.1140/epjc/s10052-024-12678-w}
}

@article{RicharteSimeone2008,
  author = {Richarte, Mart{\'i}n G. and Simeone, Claudio},
  title = {Traversable wormholes in a string cloud},
  journal = {International Journal of Modern Physics D},
  volume = {17},
  number = {8},
  pages = {1179--1196},
  year = {2008},
  doi = {10.1142/S0218271808012759},
  eprint = {0711.2297},
  archivePrefix = {arXiv},
  primaryClass = {gr-qc},
  url = {https://doi.org/10.1142/S0218271808012759}
}

@article{WaseemEtAl2025StringFluid,
  author = {Waseem, Arfa and Eid, A. and Javed, Faisal and Mustafa, G.},
  title = {Bardeen thin-shell wormholes surrounded by fluid of strings},
  journal = {Physics of the Dark Universe},
  volume = {50},
  pages = {102084},
  year = {2025},
  doi = {10.1016/j.dark.2025.102084},
  url = {https://doi.org/10.1016/j.dark.2025.102084}
}

@article{HouedokouEtAl2026HaywardLetelier,
  author = {Houedokou, Cebastien and Salako, Ines G. and Javed, Faisal and Monwanou, A. V. and Rayimbaev, Javlon},
  title = {Hayward--Letelier--AdS thin-shell wormholes with variable equations of state},
  journal = {International Journal of Modern Physics D},
  volume = {35},
  number = {4},
  pages = {2650001},
  year = {2026},
  doi = {10.1142/S021827182650001X},
  url = {https://doi.org/10.1142/S021827182650001X}
}

@article{huQEDEffectBlack2021,
  author = {Hu, Zezhou and Zhong, Zhen and Li, Peng-Cheng and Guo, Minyong and Chen, Bin},
  title = {{QED} effect on a black hole shadow},
  journal = {Physical Review D},
  volume = {103},
  number = {4},
  pages = {044057},
  year = {2021},
  doi = {10.1103/PhysRevD.103.044057},
  url = {https://doi.org/10.1103/PhysRevD.103.044057}
}

@article{grallaShapeBlackHole2020,
  author = {Gralla, Samuel E. and Lupsasca, Alexandru and Marrone, Daniel P.},
  title = {The shape of the black hole photon ring: A precise test of strong-field general relativity},
  journal = {Physical Review D},
  volume = {102},
  number = {12},
  pages = {124004},
  year = {2020},
  doi = {10.1103/PhysRevD.102.124004},
  url = {https://doi.org/10.1103/PhysRevD.102.124004}
}

@article{caoImagingSignaturesIsrael2026,
  title = {Imaging Signatures of the {{Israel}} Junction: {{Photon}} Ring Evolution in Dynamical Thin Shell {{Schwarzschild}} Spacetimes},
  shorttitle = {Imaging Signatures of the {{Israel}} Junction},
  author = {Cao, Li-Ming and Li, Long-Yue and Liu, Xia-Yuan},
  year = 2026,
  journal = {Sci. China Phys. Mech. Astron.},
  volume = {69},
  number = {5},
  pages = {250413},
  doi = {10.1007/s11433-025-2919-x}
}

@article{collaborationFirstSagittariusEvent2022,
  title = {First {{Sagittarius A}}* {{Event Horizon Telescope Results}}. {{I}}. {{The Shadow}} of the {{Supermassive Black Hole}} in the {{Center}} of the {{Milky Way}}},
  author = {Akiyama, Kazunori and others},
  collaboration = {Event Horizon Telescope},
  year = 2022,
  month = may,
  journal = {The Astrophysical Journal Letters},
  volume = {930},
  number = {2},
  pages = {L12},
  publisher = {The American Astronomical Society},
  issn = {2041-8205},
  doi = {10.3847/2041-8213/ac6674},
  urldate = {2025-12-25},
  langid = {english}
}

@article{grallaBlackHoleShadows2019,
  title = {Black {{Hole Shadows}}, {{Photon Rings}}, and {{Lensing Rings}}},
  author = {Gralla, Samuel E. and Holz, Daniel E. and Wald, Robert M.},
  year = 2019,
  journal = {Phys. Rev. D},
  volume = {100},
  number = {2},
  pages = {024018},
  doi = {10.1103/PhysRevD.100.024018}
}

@article{guerreroDoubleShadowsReflectionasymmetric2021,
  title = {Double Shadows of Reflection-Asymmetric Wormholes Supported by Positive Energy Thin-Shells},
  author = {Guerrero, Merce and Olmo, Gonzalo J. and {Rubiera-Garcia}, Diego},
  year = 2021,
  journal = {JCAP},
  volume = {2021},
  number = {04},
  pages = {066},
  doi = {10.1088/1475-7516/2021/04/066}
}

@article{heObservationalAppearanceExtra2024,
  title = {Observational Appearance and Extra Photon Rings of an Asymmetric Thin-Shell Wormhole with a {{Bardeen}} Profile},
  author = {He, Ke-Jian and Luo, Zhi and Guo, Sen and Li, Guo-Ping},
  year = 2024,
  journal = {Chin. Phys. C},
  volume = {48},
  number = {6},
  pages = {065105},
  doi = {10.1088/1674-1137/ad34bf}
}

@article{houImageKerrMelvinBlack2022,
  title = {Image of a {{Kerr-Melvin}} Black Hole with a Thin Accretion Disk},
  author = {Hou, Yehui and Zhang, Zhenyu and Yan, Haopeng and Guo, Minyong and Chen, Bin},
  year = 2022,
  journal = {Phys. Rev. D},
  volume = {106},
  number = {6},
  pages = {064058},
  doi = {10.1103/PhysRevD.106.064058}
}

@article{huQEDEffectBlack2021a,
  title = {{{QED}} Effect on a Black Hole Shadow},
  author = {Hu, Zezhou and Zhong, Zhen and Li, Peng-Cheng and Guo, Minyong and Chen, Bin},
  year = 2021,
  journal = {Phys. Rev. D},
  volume = {103},
  number = {4},
  pages = {044057},
  doi = {10.1103/PhysRevD.103.044057}
}

@article{langloisConservationLawsCollisions2002,
  title = {Conservation Laws for Collisions of Branes (or Shells) in General Relativity},
  author = {Langlois, David and Maeda, Kei-ichi and Wands, David},
  year = 2002,
  journal = {Phys. Rev. Lett.},
  volume = {88},
  pages = {181301},
  doi = {10.1103/PhysRevLett.88.181301}
}

@article{luminetImageSphericalBlack1979a,
  title = {Image of a Spherical Black Hole with Thin Accretion Disk.},
  author = {Luminet, J.-P.},
  year = 1979,
  month = may,
  journal = {Astronomy and Astrophysics},
  volume = {75},
  pages = {228--235},
  issn = {0004-6361},
  urldate = {2026-08-29},
  langid = {english}
}

@article{ohgamiWormholeShadows2015,
  title = {Wormhole Shadows},
  author = {Ohgami, Takayuki and Sakai, Nobuyuki},
  year = 2015,
  journal = {Phys. Rev. D},
  volume = {91},
  number = {12},
  pages = {124020},
  doi = {10.1103/PhysRevD.91.124020}
}

@article{olmoNewLightRings2022,
  title = {New Light Rings from Multiple Critical Curves as Observational Signatures of Black Hole Mimickers},
  author = {Olmo, Gonzalo J. and {Rubiera-Garcia}, Diego and G{\'o}mez, Diego S{\'a}ez-Chill{\'o}n},
  year = 2022,
  journal = {Phys. Lett. B},
  volume = {829},
  pages = {137045},
  doi = {10.1016/j.physletb.2022.137045}
}

@article{paulObservationalSignaturesWormholes2020,
  title = {Observational Signatures of Wormholes with Thin Accretion Disks},
  author = {Paul, Suvankar and Shaikh, Rajibul and Banerjee, Pritam and Sarkar, Tapobrata},
  year = 2020,
  journal = {JCAP},
  volume = {2020},
  number = {03},
  pages = {055},
  doi = {10.1088/1475-7516/2020/03/055}
}

@article{pengObservationalSignatureAdditional2021a,
  title = {Observational Signature and Additional Photon Rings of an Asymmetric Thin-Shell Wormhole},
  author = {Peng, Jun and Guo, Minyong and Feng, Xing-Hui},
  year = 2021,
  journal = {Phys. Rev. D},
  volume = {104},
  number = {12},
  pages = {124010},
  doi = {10.1103/PhysRevD.104.124010}
}

@article{ruedaTraversableWormholesMultiple2025,
  title = {Traversable Wormholes with Multiple Unstable Critical Curves},
  author = {Rueda, A. and Contreras, E.},
  year = 2025,
  journal = {Phys. Rev. D},
  volume = {111},
  number = {4},
  pages = {044019},
  doi = {10.1103/PhysRevD.111.044019}
}

@article{theeventhorizontelescopecollaborationFirstM87Event2019,
  title = {First {{M87 Event Horizon Telescope Results}}. {{IV}}. {{Imaging}} the {{Central Supermassive Black Hole}}},
  author = {Akiyama, Kazunori and others},
  collaboration = {Event Horizon Telescope},
  year = 2019,
  month = apr,
  journal = {The Astrophysical Journal Letters},
  volume = {875},
  number = {1},
  pages = {L4},
  publisher = {The American Astronomical Society},
  issn = {2041-8205},
  doi = {10.3847/2041-8213/ab0e85},
  urldate = {2026-01-14},
  langid = {english}
}

@article{vincentGYOTONewGeneral2011,
  title = {{{GYOTO}}: A New General Relativistic Ray-Tracing Code},
  shorttitle = {{{GYOTO}}},
  author = {Vincent, F. H. and Paumard, T. and Gourgoulhon, E. and Perrin, G.},
  year = 2011,
  journal = {Class. Quant. Grav.},
  volume = {28},
  pages = {225011},
  doi = {10.1088/0264-9381/28/22/225011}
}

@article{wangHowDifferentAre2023,
  title = {How Different Are Shadows of Compact Objects with and without Horizons?},
  author = {Wang, Xiangyu and Hou, Yehui and Guo, Minyong},
  year = 2023,
  journal = {JCAP},
  volume = {2023},
  number = {05},
  pages = {036},
  doi = {10.1088/1475-7516/2023/05/036}
}

@article{wangNovelShadowsAsymmetric2020,
  title = {Novel Shadows from the Asymmetric Thin-Shell Wormhole},
  author = {Wang, Xiaobao and Li, Peng-Cheng and Zhang, Cheng-Yong and Guo, Minyong},
  year = 2020,
  month = dec,
  journal = {Physics Letters B},
  volume = {811},
  eprint = {2007.03327},
  primaryclass = {gr-qc},
  pages = {135930},
  issn = {03702693},
  doi = {10.1016/j.physletb.2020.135930},
  urldate = {2026-08-26},
  archiveprefix = {arXiv}
}

@article{wielgusReflectionasymmetricWormholesTheir2020,
  title = {Reflection-Asymmetric Wormholes and Their Double Shadows},
  author = {Wielgus, Maciek and Horak, Jiri and Vincent, Frederic and Abramowicz, Marek},
  year = 2020,
  journal = {Phys. Rev. D},
  volume = {102},
  number = {8},
  pages = {084044},
  doi = {10.1103/PhysRevD.102.084044}
}

@article{zengQEDAccretionFlow2022,
  title = {{{QED}} and Accretion Flow Models Effect on Optical Appearance of {{Euler}}--{{Heisenberg}} Black Holes},
  author = {Zeng, Xiao-Xiong and He, Ke-Jian and Li, Guo-Ping and Liang, En-Wei and Guo, Sen},
  year = 2022,
  journal = {Eur. Phys. J. C},
  volume = {82},
  number = {8},
  pages = {764},
  doi = {10.1140/epjc/s10052-022-10733-y}
}

@article{zhangImagingThickAccretion2024a,
  title = {Imaging Thick Accretion Disks and Jets Surrounding Black Holes},
  author = {Zhang, Zhenyu and Hou, Yehui and Guo, Minyong and Chen, Bin},
  year = 2024,
  journal = {JCAP},
  volume = {2024},
  number = {05},
  pages = {032},
  doi = {10.1088/1475-7516/2024/05/032}
}

\end{document}